\documentclass[epj,nopacs]{svjour}

\usepackage{amssymb}
\usepackage{epsfig}
\usepackage{eucal}
\usepackage{amsmath}

\newcommand{\beq}{\begin{equation}} 
\newcommand{\eeq}{\end{equation}}   
\newcommand{\bea}{\begin{eqnarray}} 
\newcommand{\eea}{\end{eqnarray}}
\newcommand{\non}{\nonumber}
         
\newcommand{\Eq}[1]{Eq.~(\ref{#1})}

\def\Li{\hbox{Li}}  
\def\DAF{DA\char8NE}                                  

\headnote{\hfill \parbox{4cm}{\vspace{-2cm}
\tt \normalsize CERN-TH/2003-179 \\ TTP03-22}}

\title{The radiative return at $\boldsymbol{\Phi}$- and $\boldsymbol{B}$-factories: \\
 FSR at next-to-leading order
\thanks{Work supported in part by BMBF under grant number 05HT9VKB0,
EC 5th Framework Programme under contracts HPRN-CT-2000-00149,
and HPRN-CT-2002-00311 (EURIDICE network),
KBN under contract 2 P03B 017 24, 
Generalitat Valenciana under grant CTIDIB/2002/24, 
and MCyT under grants FPA-2001-3031 and BFM2002-00568.}}

\author{
Henryk Czy\.z\inst{1}
\thanks{\email{czyz@us.edu.pl}} 
\and
Agnieszka Grzeli{\'n}ska\inst{1}
\thanks{\email{grzel@server.phys.us.edu.pl}; 
Supported in part by `Marie Curie Training Site' at Universit\"at Karlsruhe.}
\and
Johann H. K\"uhn\inst{2}
\thanks{\email{johann.kuehn@uni-karlsruhe.de}}
\and
Germ\'an Rodrigo\inst{3}
\thanks{\email{german.rodrigo@cern.ch}; 
Supported by EC 5th Framework Programme under contract HPMF-CT-2000-00989.}
}

\institute{
Institute of Physics, University of Silesia,
PL-40007 Katowice, Poland. \and
Institut f\"ur Theoretische Teilchenphysik,
Universit\"at Karlsruhe, D-76128 Karlsruhe, Germany.\and
Theory Division, CERN, CH-1211 Geneva 23, Switzerland.}

\date{Received: August 29, 2003}

\abstract{ The measurement of the pion form factor and, more
generally, of the cross section for electron--positron annihilation
into hadrons through the radiative return has become an important
task for high luminosity colliders such as the $\Phi$- or
$B$-meson factories. For detailed understanding and analysis
of this reaction, the construction of a Monte Carlo program, 
PHOKHARA, has been undertaken. Version 2.0 was based on a 
next-to-leading order (NLO) treatment of the corrections from initial-state 
radiation (ISR). In the present paper a further extension of
PHOKHARA (version 3.0) is described, which incorporates NLO
corrections to final-state radiation (FSR). The impact of combined 
ISR and FSR on various distributions is investigated and methods 
are presented, which will allow the extraction of the form factor,
and even give access to inclusive photon emission due to FSR. 
The dependence of the results on the model for FSR is discussed and 
the impact of this contribution on the anomalous magnetic moment of 
the muon is evaluated.}

\begin{document}


\maketitle


\section{Introduction}

Electron--positron annihilation into hadrons is one of the basic
reactions of particle physics, crucial for the understanding
of hadronic interactions. At high energies, around the $Z$ resonance,
the measurement of the inclusive cross section and its interpretation
within perturbative QCD~\cite{CKK,HS} give rise to one of the most
precise and theoretically founded  determinations of the strong coupling
constant $\alpha_s$~\cite{EWG}. Also, measurements in the intermediate 
energy region, between 3 GeV and 11 GeV can be used to determine $\alpha_s$
and at the same time give rise to precise measurements of charm 
and bottom quark masses~\cite{KS}. The low energy region is crucial
for predictions of the hadronic contributions to $a_\mu$, the anomalous 
magnetic moment of the muon, and to the running of the electromagnetic
coupling from its value at low energy up to $M_Z$ 
(for reviews see e.g.~\cite{a_mu1,Jegetc.,Melnikov:2001uw,Davier:2002dy,HMNT02};
the most recent experimental result for $a_\mu$ is presented in~\cite{Bennet}).
Last, but not least, the investigation of the exclusive final states
at large momenta allows for tests of our theoretical 
understanding of form factors within the framework of perturbative QCD.
Beyond the intrinsic interest in this reaction,
these studies may provide important clues for the interpretation
of exclusive decays of B-mesons, a topic of evident importance for
the extraction of CKM matrix elements. 

Measurements of the cross section for electron--positron annihilation
were traditionally performed in the scanning mode, i.e. by varying the 
beam energies of the collider. The recent advent of 
$\Phi$- and $B$-meson factories with enormous luminosities
allows us to exploit the radiative return to explore the whole energy
region from threshold up to the nominal energy of the collider. 
Photon radiation from the initial state reduces the cross section by a factor
${\cal O}(\alpha/\pi)$. However, this is easily compensated by the 
enormous luminosity of `factories' and the advantage of performing
the measurement over a wide range of energies in one homogeneous data
sample~\cite{Binner:1999bt} (for an early proposal along these lines, 
see~\cite{Zerwas}). In principle the reaction 
$e^+e^- \to \gamma + {\mathrm {hadrons}}$ receives contributions from 
both initial- and final-state radiation. Only the former is of interest
for the radiative return; the latter has to be eliminated by
suitably chosen cuts. The proper analysis thus requires the construction
of Monte Carlo event generators.
The event generator EVA was based on a leading order treatment of ISR and FSR, 
supplemented by an approximate inclusion of additional collinear radiation 
based on structure functions and included two-pion~\cite{Binner:1999bt} and 
four-pion final states~\cite{Czyz:2000wh}.
Subsequently the event generator PHOKHARA was developed; it is based
on a complete next-to-leading order (NLO) treatment of 
ISR~\cite{Rodrigo:2001jr,Rodrigo:2001cc,Rodrigo:2001kf,Kuhn:2002xg,Czyz:2002np,Kuhn:2001,Rodrigo:2002hk}.
In its version 2.0 it included ISR at NLO and FSR at leading order (LO)
for $\pi^+ \pi^-$ and $\mu^+ \mu^-$ final states (and four-pion
final states without FSR in the formulation described in detail 
in~\cite{Czyz:2000wh}).

Recent preliminary experimental results indeed demonstrate the power
of the method and seem to indicate that a precision of one per cent
or better is within reach~\cite{CDKMV2000,Adinolfi:2000fv,Aloisio:2001xq,Denig:2001ra,babar,Barbara:Morion,Berger:2002mg,Venanzoni:2002,Achim:radcor02,KLOE:2003,Blinov,Juliet}.
In view of this progress a further improvement of our theoretical 
understanding seems to be required. In the present work we therefore present 
a new version (version 3.0) of PHOKHARA, 
which allows for the `simultaneous' emission of one photon from 
the initial and one photon from the final state, requiring only one
of them to be hard. This includes in particular the radiative return
to $\pi^+ \pi^- (\gamma)$ and thus the measurement of the (one-photon) 
inclusive $\pi^+ \pi^-$ cross section. The new
version of PHOKHARA and various tests of its technical precision
will be presented and several suggestions for a model-independent
extraction of the form factor will be discussed.

The issue of photon radiation from the final states is closely 
connected to the question of $\pi^+ \pi^- (\gamma)$ 
contributions to $a_{\mu}$ (for related discussions, 
see e.g.~\cite{Melnikov:2001uw,Gluza:rad}).
Soft photon emission is clearly described by
the point-like pion model. Hard photon emission, with 
$E_{\gamma} \geq {\cal O}$(100 MeV), however, might be sensitive
to unknown hadronic physics. We will therefore study the size of
virtual, soft and hard corrections separately, and argue that 
contributions from the hard region, above 100 MeV, are small with respect 
to the present experimental and theory-induced uncertainty.

The outline of the present paper is the following. 
In section 2 we recall the basic aspects of the radiative return
in leading order, define strategies to separate ISR and FSR
through cuts, and discuss their markedly different angular distribution 
and the characteristic feature of ISR--FSR interference.
In section 3 we present the estimate of $\pi^+\pi^-(\gamma)$ 
contributions to $a_{\mu}$. Section 4 is devoted to the description
of the additional two-photon corrections originating from
simultaneous ISR--FSR emission and their implementation in PHOKHARA 3.0.
The impact of these new terms on various distributions is described
in section 5. A brief summary and our conclusions can be found in
section 6.

\section{ISR versus FSR and the radiative return at leading order}
\label{sec:return}

\begin{figure}[h]
\begin{center}
\epsfig{file=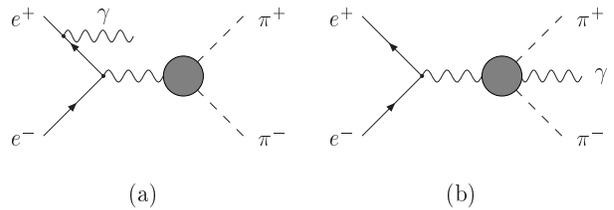,width=8.5cm} 
\caption{Leading order contributions to the reaction
$e^+e^-\to\pi^+\pi^-\gamma$ from ISR (a) and FSR (b).}
\label{fig1}
\end{center}
\end{figure}

Let  us, in  a first  step, recall  the basic  ingredients for  a proper
exploitation  of  the  enormous  luminosity  at  $\Phi$-  and  $B$-meson
factories  through  the  radiative   return.   In  this  paper  we  will
concentrate  on  the  measurement   of  the  $\pi^+\pi^-$  final  state,
accompanied by one or two photons.   We shall start with a discussion of
the leading process,  with only one photon, radiated  from the electron
(positron) or the hadronic system:
\begin{equation}
e^+ \ e^- \to \pi^+(p^+) \  \pi^-(p^-) \ \gamma~. 
\end{equation}
For the radiative return this corresponds to the Born approximation, and
higher  order corrections  will  be discussed  below.  The amplitude  of
interest  describes  radiation  from  the  initial state  (ISR)  and  is
depicted  in  Fig.~\ref{fig1}a (permutations  of  photon  lines  will always  
be omitted). It is  proportional  to the  pion  form factor, evaluated  at
$Q^2=(p^+  + p^-)^2$. However,  radiation from  the charged  pions (FSR)
(Fig.~\ref{fig1}b) evidently leads to the same final state and must be suppressed
and controlled with sufficient precision.  A variety of methods, which have
already been  described in~\cite{Binner:1999bt,Rodrigo:2001kf,Czyz:2002np}, 
will be  recalled in the following.

The fully differential cross section describing photon emission can 
be split into three pieces
\begin{equation}
d\sigma = d\sigma_\mathrm{ISR} + d\sigma_\mathrm{FSR} + d\sigma_\mathrm{INT}~,
\end{equation}
which originate  from the  squared ISR and  FSR amplitudes respectively,
plus the  interference term. They  depend on two Dalitz  variables, which
characterize the energies  of $\pi^+$ and $\pi^-$ and  the photon, and on
the   three   Euler   angles   describing   the   orientation   of   the
$\pi^+\pi^-\gamma$ production plane in  the centre-of-mass system (cms).
The  dependence on  the azimuthal  angle  around the  beam direction  is
trivial in the case of unpolarized beams.

The  interference  term, $d\sigma_\mathrm{INT}$,  is  antisymmetric under
the exchange of $\pi^+$ and $\pi^-$ or  $e^+$ and $e^-$. This allows us to test
the model for FSR  and, furthermore, can even lead to a measurement of
the amplitude $A_\mathrm{FSR}$. The usefulness of this method has already 
been emphasized in~\cite{Binner:1999bt}.

\begin{figure*}[ht]
{\hskip-.5cm\epsfig{file=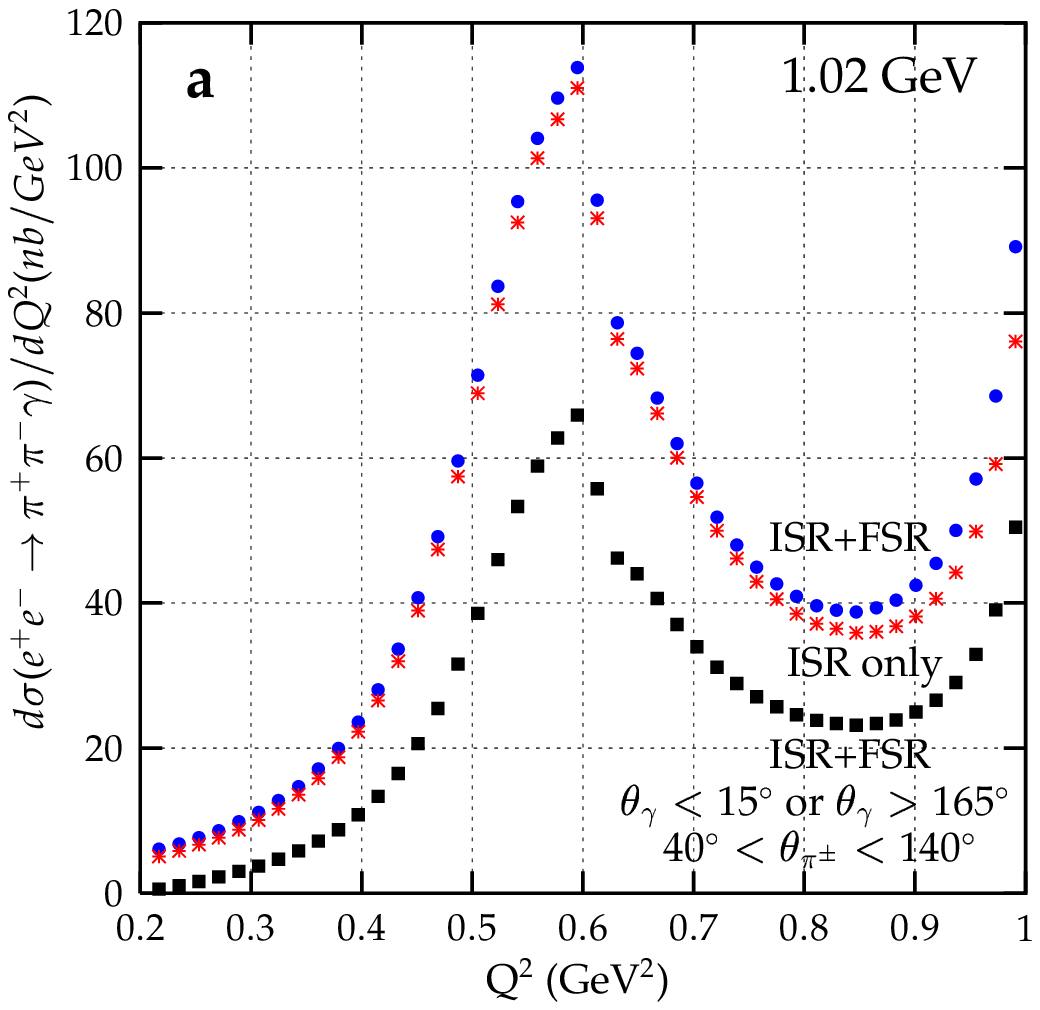,height=7cm,width=9.4cm} 
\hskip-0.3cm\epsfig{file=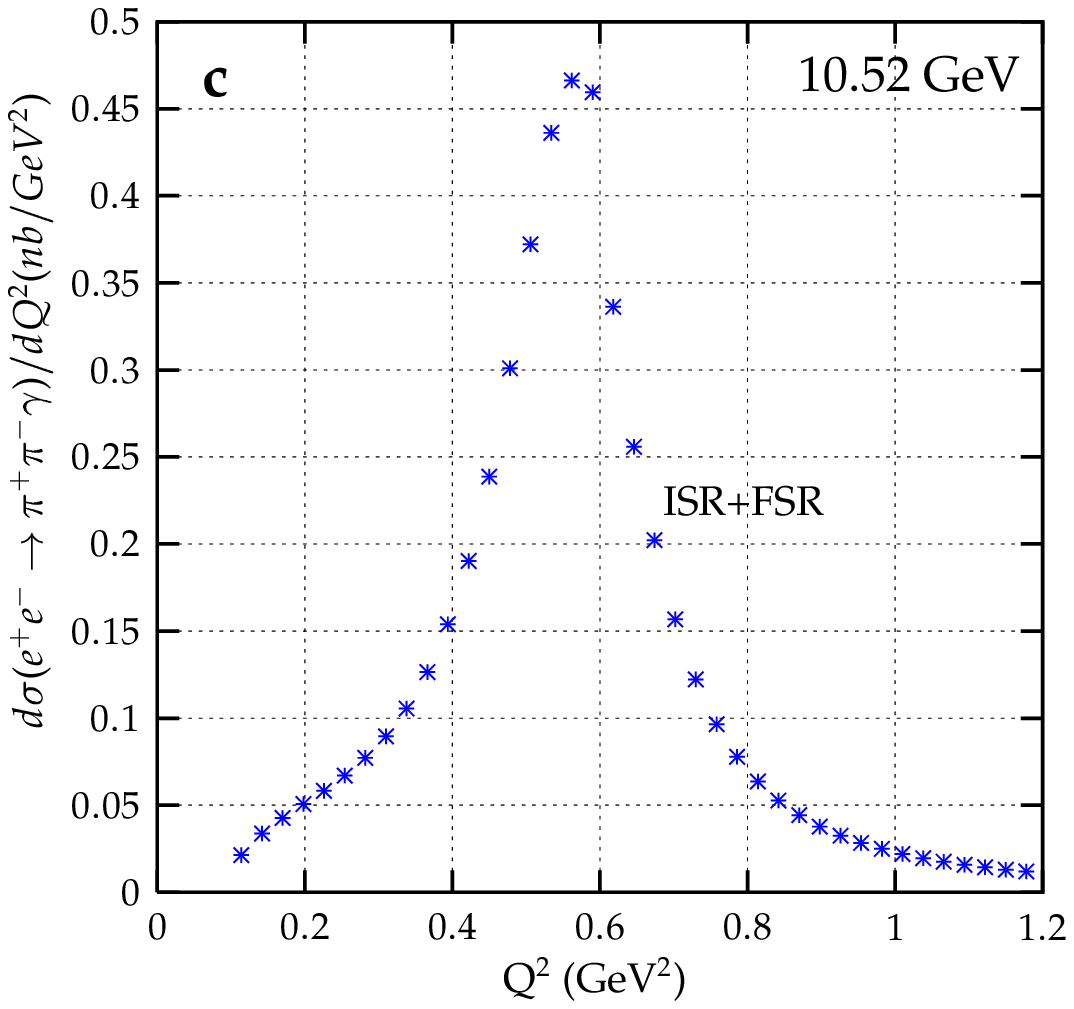,height=7cm,width=9.4cm} } 
{\phantom{}\hskip-0.5cm\epsfig{file=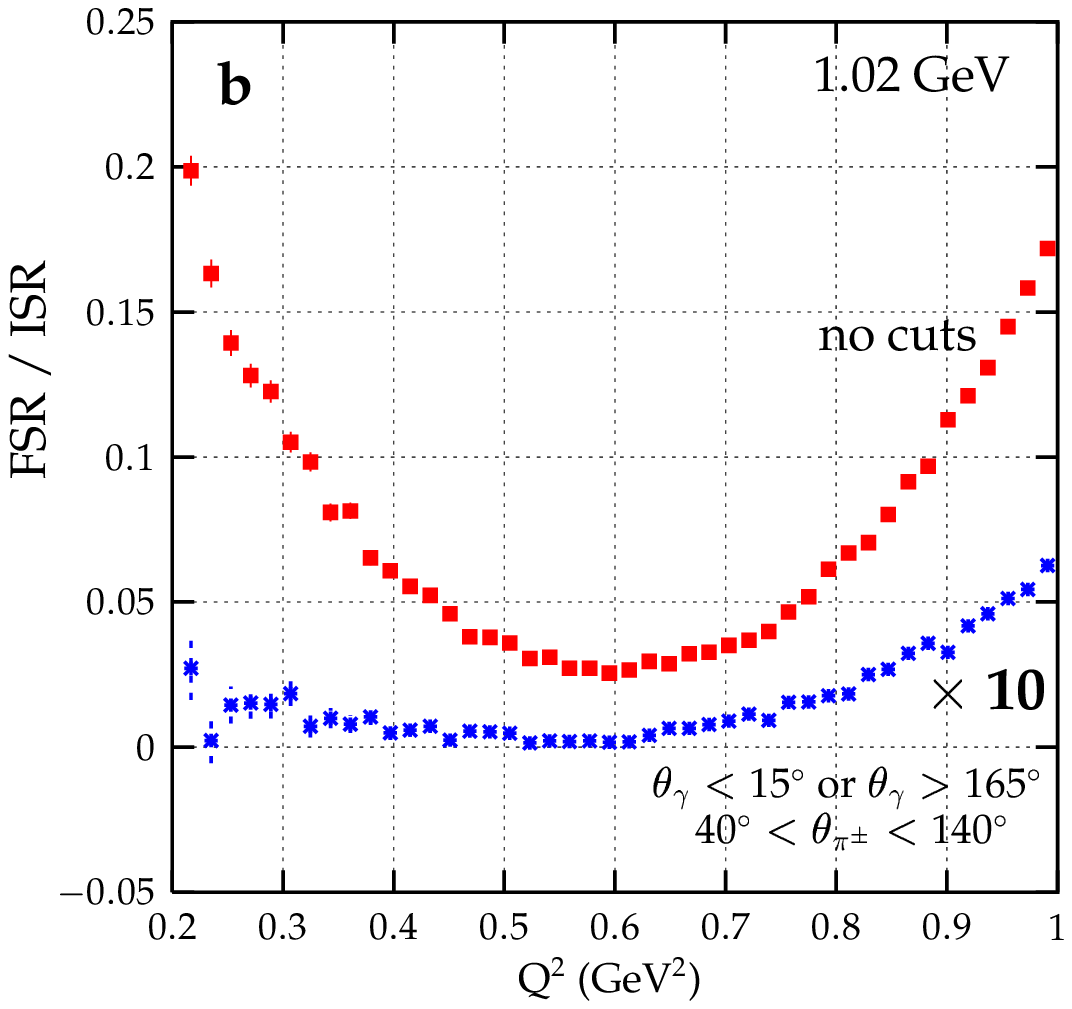,height=7cm,width=9.4cm} 
\hskip-0.3cm\epsfig{file=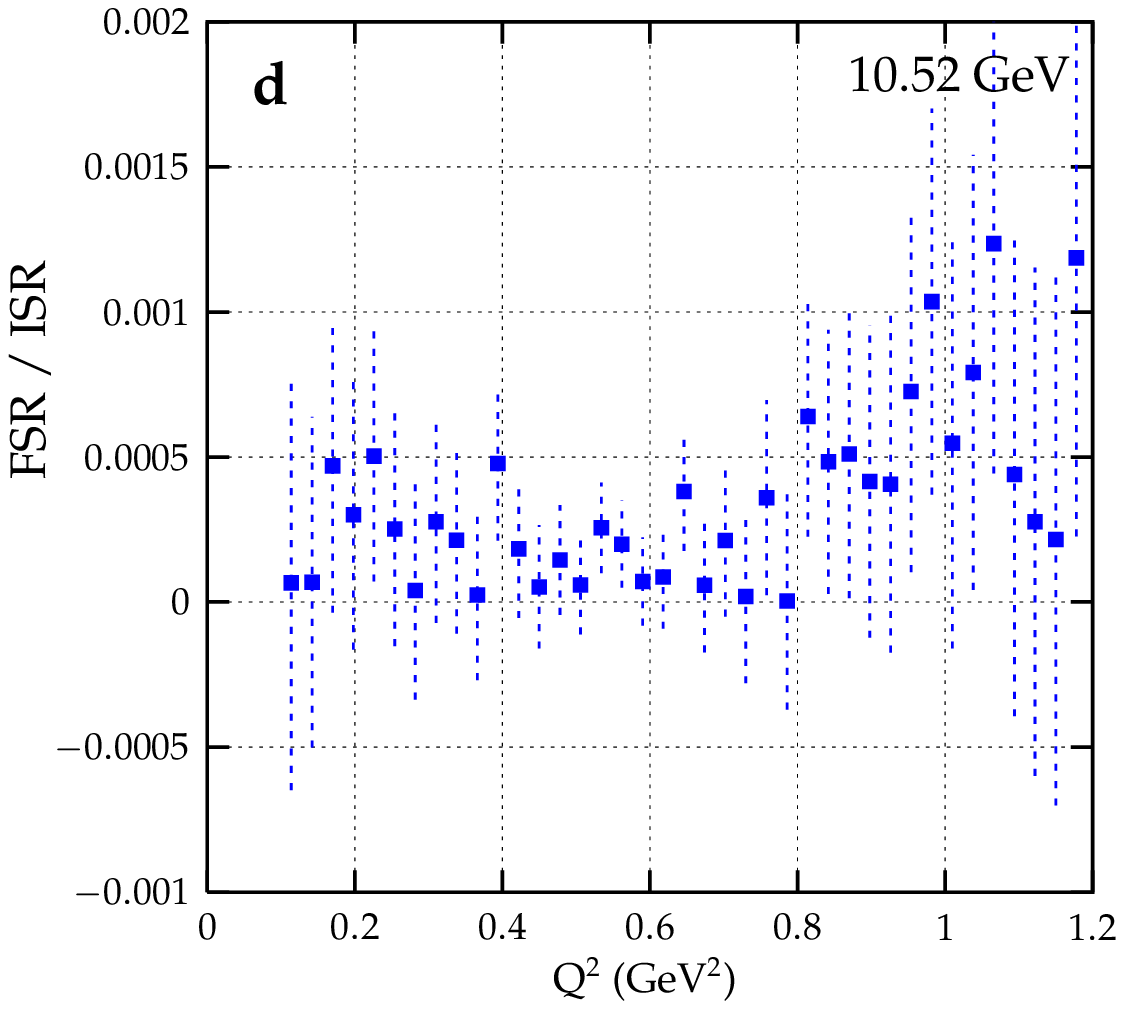,height=7cm,width=9.4cm} } 
\caption{Contributions to the inclusive photon spectrum from ISR 
compared to ISR plus FSR at \(\sqrt{s}~=\) 1.02 GeV (a) (without and with 
cuts) and  \(\sqrt{s}~=\) 10.52 GeV (c). In (b) and (d) the relative contribution
of FSR is plotted (multiplied by a factor of 10 for \(\sqrt{s}~=\) 1.02 GeV, 
when cuts are applied). }
\label{fig2}
\end{figure*}

\begin{figure}[ht]
\epsfig{file=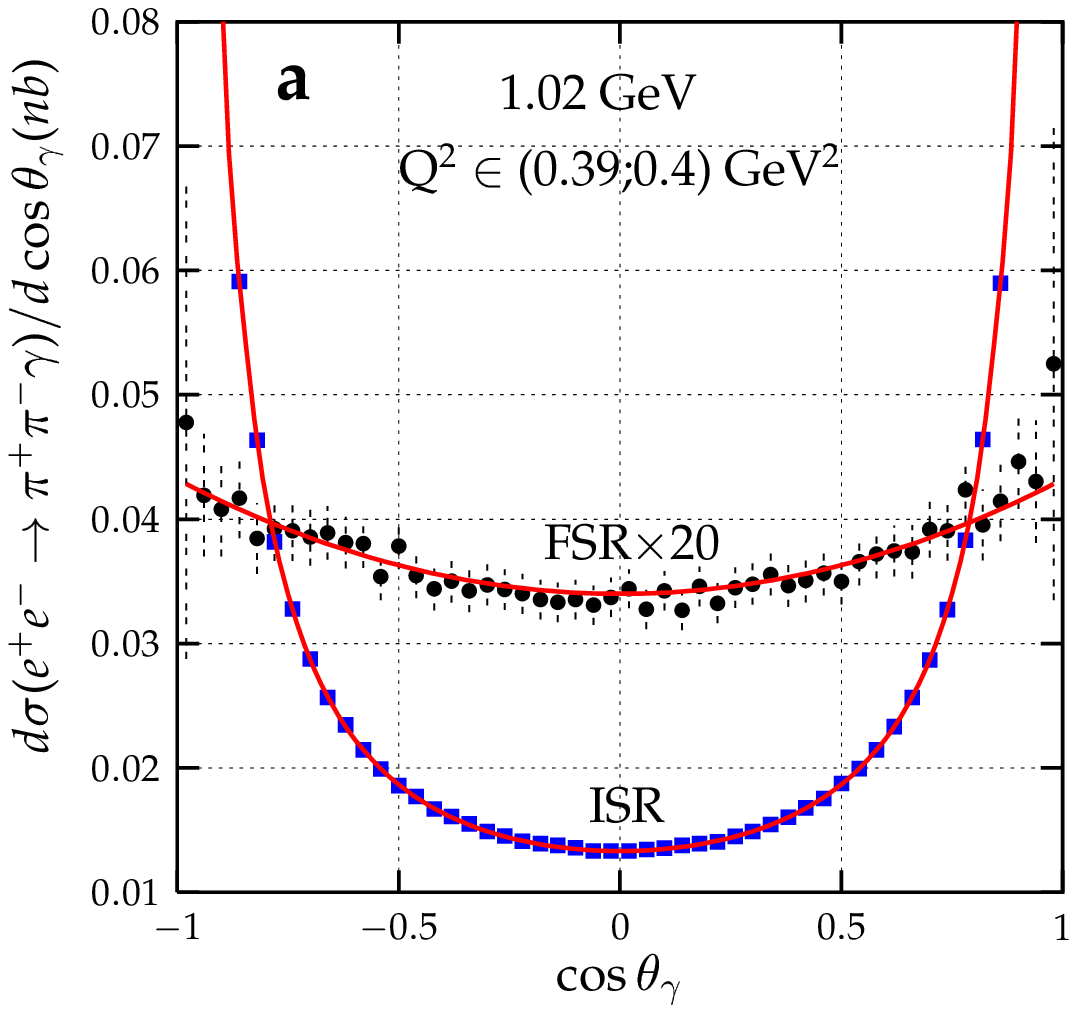,height=5.9cm,width=8.8cm}
\epsfig{file=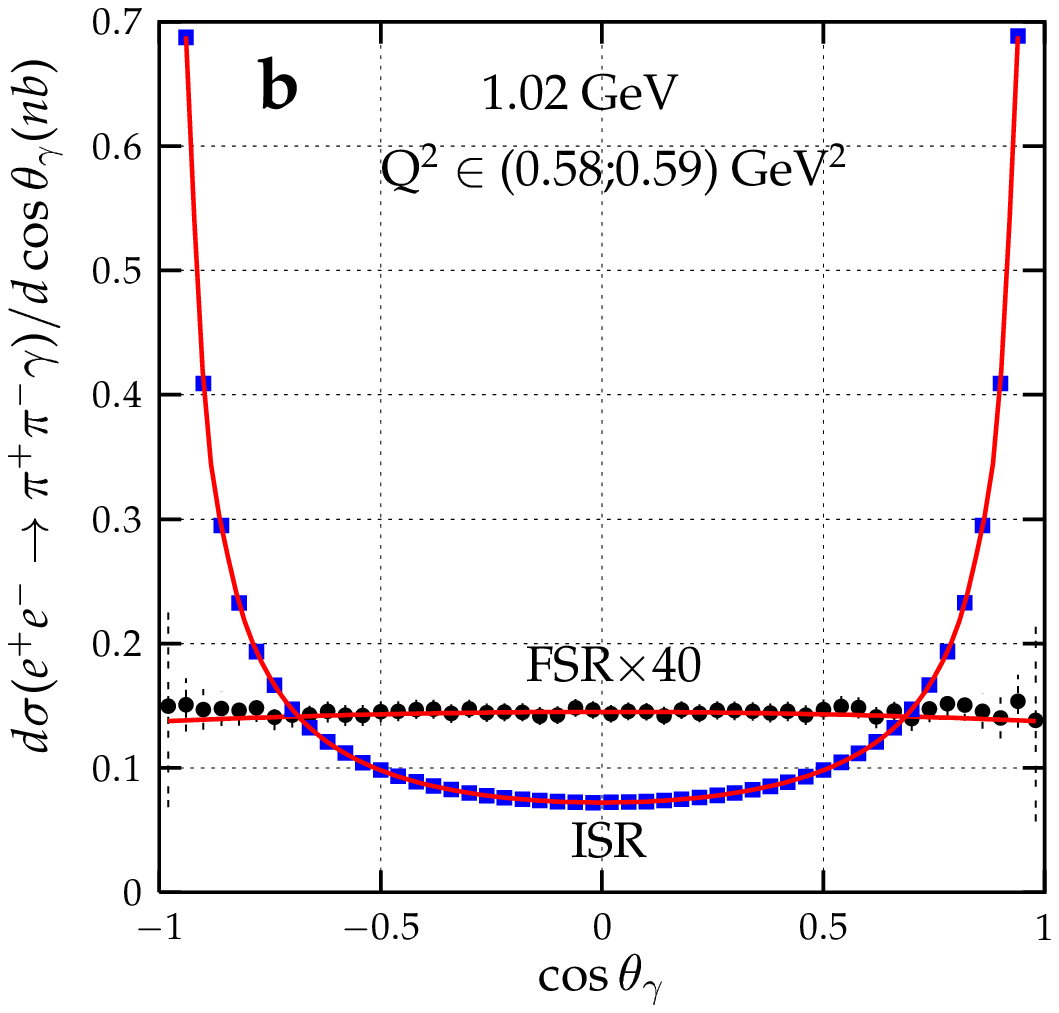,height=5.9cm,width=8.8cm}
\epsfig{file=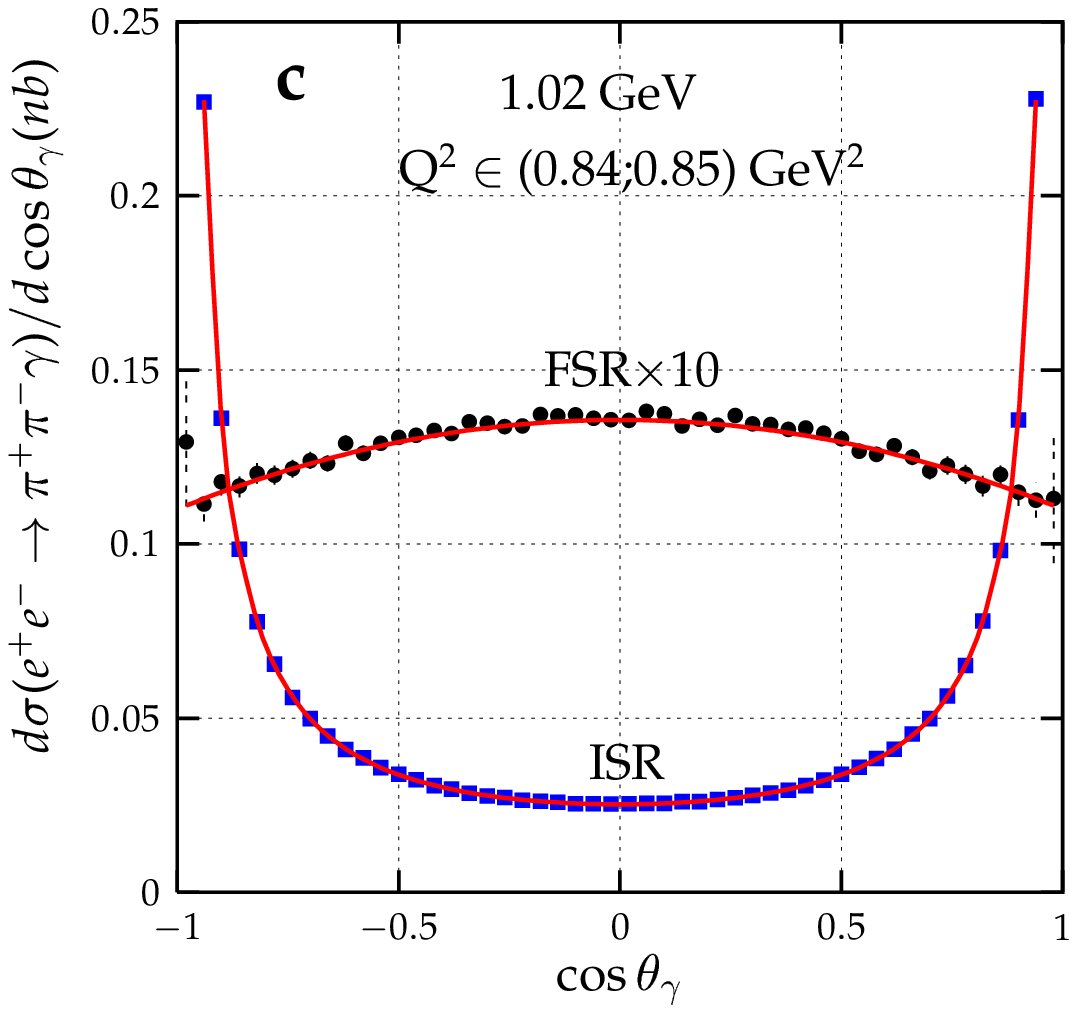,height=5.9cm,width=8.8cm}
\begin{center}
\caption{Angular distribution of photons with $E_\gamma \in (314,319)$ MeV (a), 
$E_\gamma \in (221,226)$~MeV (b) and $E_\gamma \in (93,98)$~MeV (c), separated 
according to ISR and FSR. The curves represent the
fits according to \Eq{eq:angdistr}.}
\label{fig3}
\end{center}
\end{figure}

Let us, for the moment, concentrate on the charge-symmetric pieces. The
inclusive photon spectrum, separated according to ISR and  FSR, is shown
in  Fig.~\ref{fig2}. The discrimination  between  ISR and  FSR is based on the
markedly different angular distribution  of  the  photon. Without any
assumption on the nature of FSR, the cross section can be written in the 
form
\begin{eqnarray}
Q^2 \frac{d\sigma}{d Q^2 d\cos\theta_\gamma} &=&
\frac{4\alpha^3}{3s} |F(Q^2)|^2 \frac{\beta^3(s)}{4} \non \\ &\times&
\bigg[ \frac{(s^2 + Q^4)}{s(s-Q^2)}\frac{1}{1-\cos^2\theta_\gamma} 
-\frac{s-Q^2}{2s}\bigg] \non \\ 
 &\kern -35pt+&\kern -20pt
  A_1(s,Q^2)\cos^2\theta_\gamma + A_2(s,Q^2)\sin^2\theta_\gamma~,
\label{eq:angdistr}
\end{eqnarray}
where $Q^2$ is the invariant mass of the $\pi^+,\pi^-$ system, $\theta_\gamma$
is the photon polar angle and $\beta(s) = \sqrt{1-4m_\pi^2/s}$.
Terms  $\sim m_e^2/s$ are  neglected for  simplicity, while they are
present in the program.  Fitting the observed  photon distribution to
this form allows an unambiguously separation of ISR and  FSR. The composition
of  the angular  distribution  of  photons from  the  two components  is
displayed in Fig.~\ref{fig3} for three photon energies.

The second option is based on  the fact that FSR is dominated by photons
collinear to  $\pi^+$ or  $\pi^-$, ISR by  photons collinear to  the beam
direction.  This suggests  that we should consider only  events with photons  
well separated from  the charged pions  and preferentially close to  the 
beam~\cite{Binner:1999bt,Rodrigo:2001kf,Czyz:2002np}. 
The contamination of the events with FSR is reduced to  
less than five per mille (Fig.~\ref{fig2}b). Considering just the
inclusive photon distribution, without inclusion of any cut would 
necessarily have led us to wrong conclusions. For higher beam energies,
the LO FSR is naturally suppressed and at $\sqrt{s}$ = 10.52 GeV it 
is entirely negligible  (Fig.~\ref{fig2}d). 

\begin{figure}[ht]
\epsfig{file=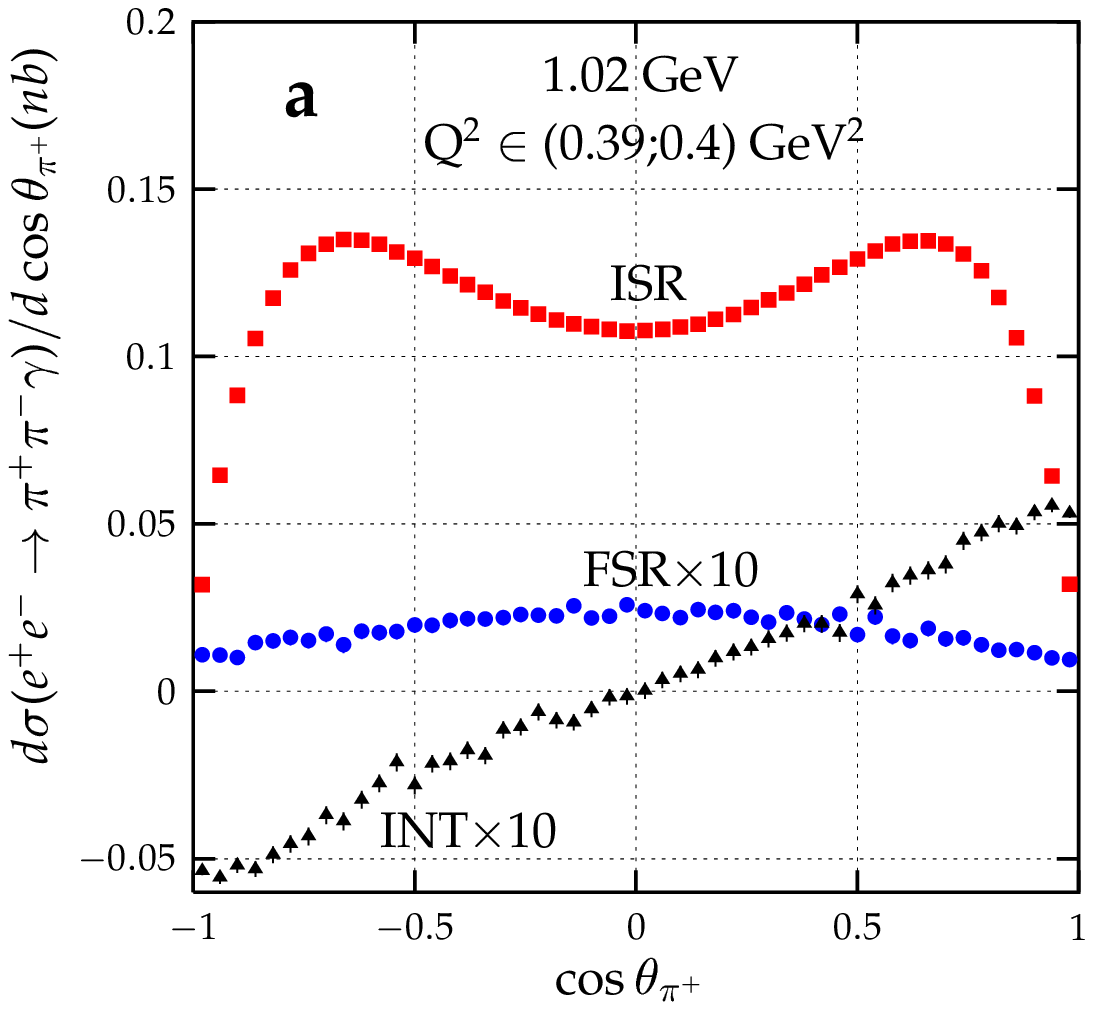,height=5.9cm,width=8.8cm}
\epsfig{file=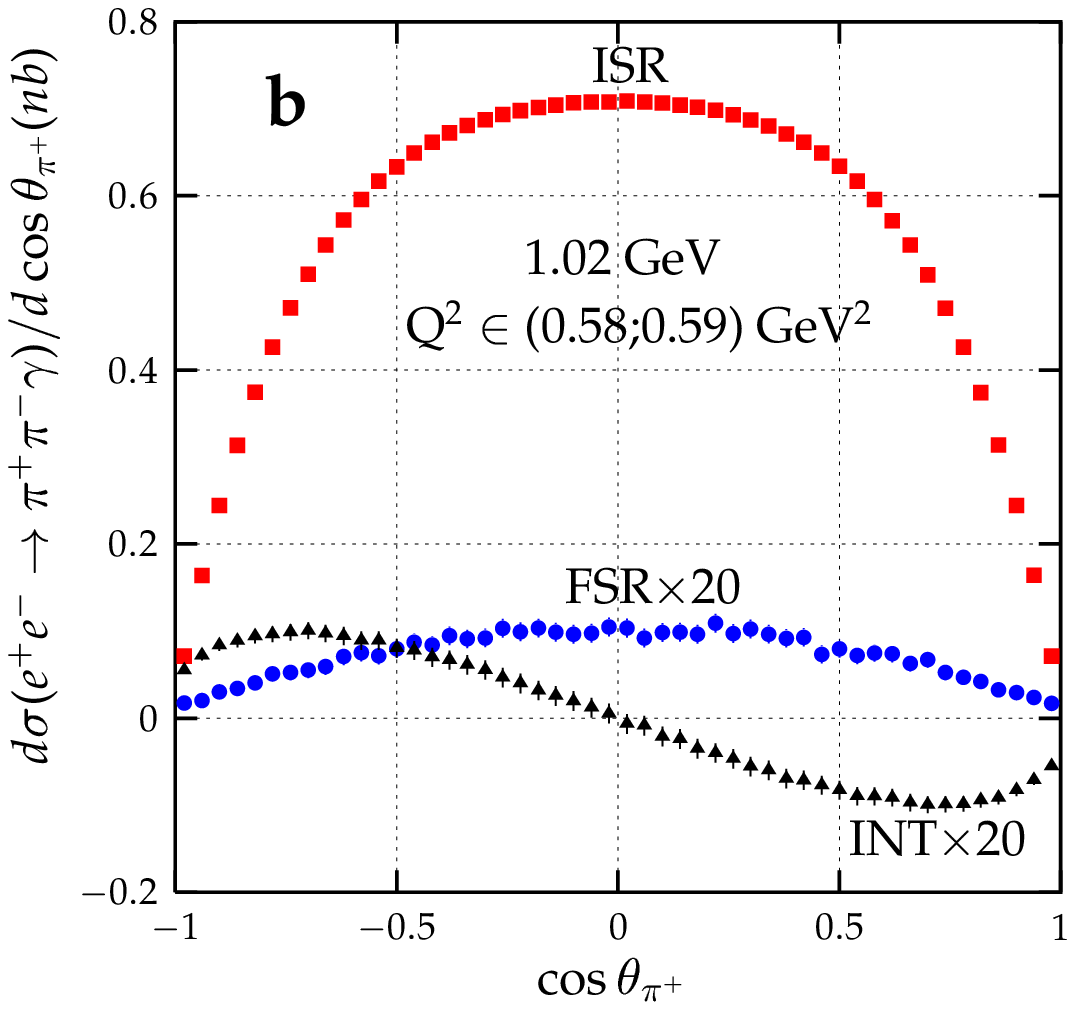,height=5.9cm,width=8.8cm}
\epsfig{file=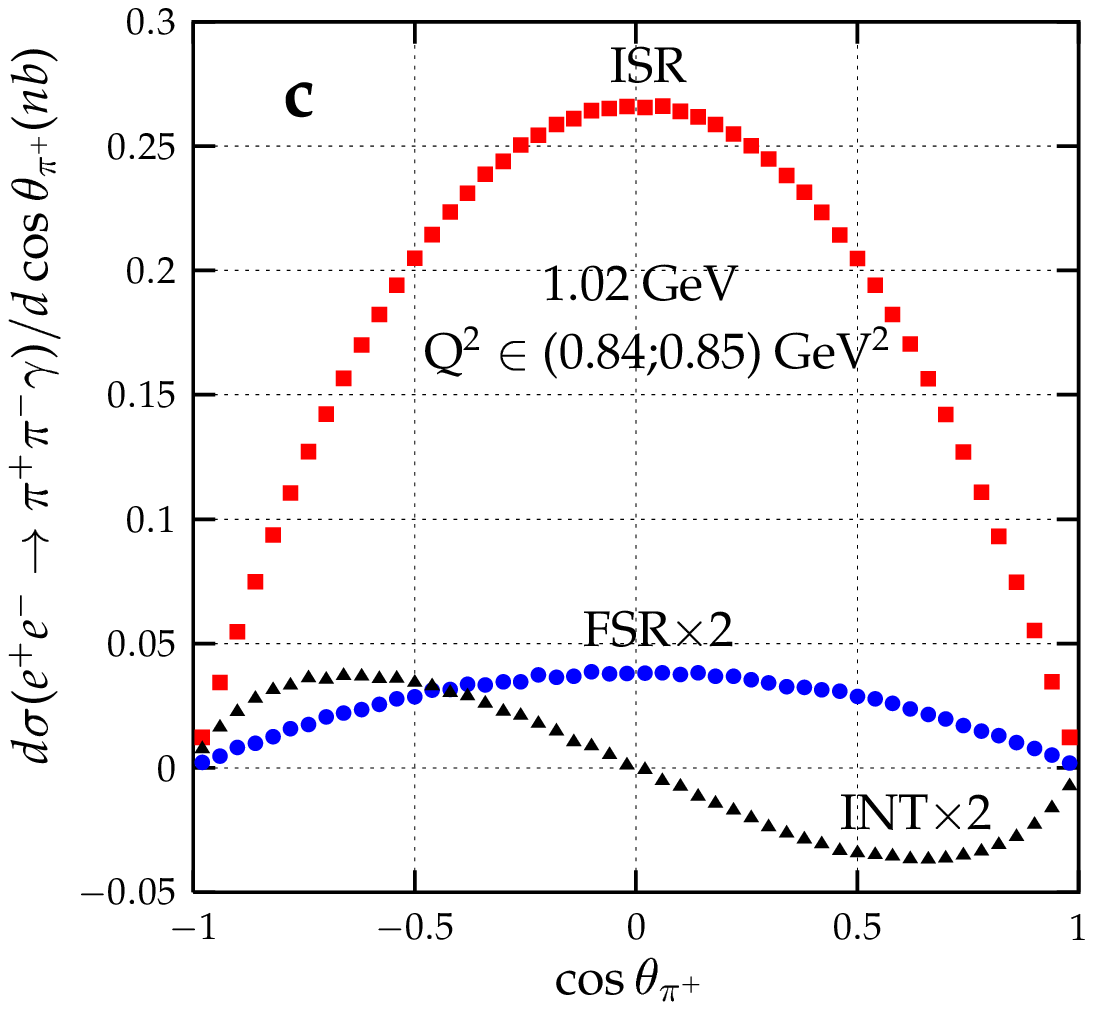,height=5.9cm,width=8.8cm}
\begin{center}
\caption{Contributions of ISR, FSR and INT terms to
pion polar angle differential cross section for three different
ranges of \(Q^2\). }
\label{fig4}
\end{center}
\end{figure} 

The interference term is asymmetric under the exchange of $\pi^+$ and 
$\pi^-$ or electron and positron:
\begin{equation}
d\sigma_\mathrm{INT}(p^+,p^-) = -d\sigma_\mathrm{INT}(p^-,p^+). 
\end{equation}
This gives  rise to a  forward--backward asymmetry of the  inclusive pion
distribution~\cite{Binner:1999bt}. 
In kinematical regions where FSR is relatively more pronounced,
the interference term is relatively large
(Fig.~\ref{fig4}c, \(Q^2\) between 0.84 and 0.85~GeV$^2$)
and vice versa (Figs.~\ref{fig4}a and~\ref{fig4}b). Note that all these
considerations refer to the low energy analysis of KLOE. At high
energies, e.g. at  $B$-factories, ISR and FSR are clearly separate 
for the $\pi^+\pi^-\gamma$ final state. 
 
Similar tests were already performed by KLOE~\cite{Aloisio:2001xq}, where
it was demonstrated that inclusive photon asymmetries are in excellent
agreement with the EVA MC~\cite{Binner:1999bt}, where FSR emission is
modelled by point-like pions. Given enough
statistics, the charge-asymmetric piece of the cross section can even be
investigated for every point in the Dalitz plane, e.g. as a function of
$Q^2$  and the angle between $\pi^+$ and $\gamma$ separately. Keeping
these variables fixed
and varying the angle between photon and electron, the relative amount
of the ISR amplitude can be varied drastically, while the FSR amplitude
stays roughly constant. 

Alternatively, we could formulate the angular distributions in
terms of independent helicity amplitudes for FSR, which depend on
the two Dalitz variables only, and deduce information on these
amplitudes from their interference with the ISR amplitude. In the
absence of any model for FSR beyond scalar QED (sQED) 
and in view of the fact that sQED will presumably provide a 
satisfactory description at \DAF \  energies we
shall not dwell further on this subject.

Let us summarise the main points of this `leading order' discussion:
\begin{itemize}
\item[1.]{The cross sections for photon emission from 
   ISR and FSR can be disentangled as a consequence of the marked 
   difference in angular distributions between the two processes, 
   \Eq{eq:angdistr}. This observation is completely general
   and does not rely on any model like sQED for FSR. This
   allows us to measure the cross section for
   $\sigma(e^+e^-\to\gamma^*\to\pi^+\pi^-\gamma)$ directly for fixed $s$ as
   a function of $E_\gamma$, an important ingredient in the analysis of
   hadronic contributions to $a_\mu$, as discussed in
   Section~\ref{sec:amu}. Alternatively we can employ suitable angular 
   cuts to separate the two components.}

\item[2.]{Various charge-asymmetric distributions can be used for independent
   tests of the FSR model amplitude, with the forward--backward
   asymmetry as simplest example. Typically the charge asymmetry is
   large in the region where ISR  and FSR are comparable in size and
   small when the separation is clean and simple from general
   considerations. A typical case for this second possibility is the
   radiative return at $\sqrt{s} = 10.52$~GeV, at the $B$-meson
   factories, where the $\pi^+\pi^-\gamma$ final state is completely
   dominated by ISR.}
\end{itemize}

\section{Higher order contributions to the anomalous magnetic moment of the muon}
\label{sec:amu}

In leading order the invariant mass of the $\pi^+\pi^-\gamma$ system
is restricted to the cms energy $\sqrt{s}$. If the relevant angular
distributions, as predicted by sQED, coincide with the
experimental analysis, the validity of this model at lower energies is
highly plausible. Nevertheless, a measurement of $\gamma^*\to \pi^+\pi^-\gamma$
for variable $\sqrt{s}$ is desirable for an independent cross check. In
Section~\ref{sec:fsrnlo} we will demonstrate that this is indeed possible with the
radiative return, if events with two photons in  the final state are
investigated. This is in fact one of the motivations for extending the
event generator PHO\-KHA\-RA to events with simultaneous ISR and FSR. However,
before discussing this issue in detail, an investigation of
$\gamma^*\to \pi^+\pi^-\gamma$, the corresponding virtual corrections
and the relevance of these amplitudes to the analysis of the muon
anomalous magnetic moment $a_\mu$ is in order.

Let us concentrate on the one-particle irreducible ha\-dro\-nic
contributions
\begin{equation}
a_\mu^\mathrm{had,LO} = 
\frac{\alpha^2}{3 \pi^2} \int_{4m_{\pi}^2}^{\infty}  
\frac{ds}{s} \; K(s) \; R(s) \ \ ,
\end{equation}
with the familiar kernel $K$ and the $R$ ratio defined through
\begin{equation}
R(s)\propto|\langle0|J_\mu|\mathrm{had},(\gamma)\rangle|^2.
\end{equation}
$R(s)$ can be obtained from the cross section for electron--positron
annihilation into hadrons, after correcting for ISR
and for the modifications of the photon propagator, which are
sometimes (not quite correctly in the time-like region) summarized as
`running' of the fine structure constant.

At the present level of precision of hadronic contributions to
\(a_\mu\), roughly half to one per cent, photonic corrections to 
final states with hadrons start to become relevant. 
Qualitatively the order of magnitude of this effect can be
estimated either by using the quark model with $m_u\approx m_d\approx
m_s \approx 180$~MeV (adopted to describe the lowest order 
contribution~\cite{piv}), or with  $m_u\approx m_d\approx m_s \approx 66$~MeV
(adopted to describe the lowest order contribution to $\alpha(M_Z)$) 
or by using $\pi^+\pi^-$ as dominant intermediate hadronic state plus 
photons coupled according to sQED. The three estimates 
(see also~\cite{Melnikov:2001uw})
\bea
\delta a_\mu({\rm quark},\gamma,m_q=180~{\rm MeV})
 = \ 1.880 \times 10^{-10}~, \nonumber \\
\delta a_\mu({\rm quark},\gamma,m_q=66~{\rm MeV})
 = \ 8.577 \times 10^{-10}~, \nonumber \\
 \delta a_\mu(\pi^+\pi^-,\gamma) = \ 4.309 \times 10^{-10}~,
\eea
are comparable in magnitude and begin to be relevant at the present
level of precision. This order-of-magnitude estimate
suggests that a more careful analysis is desirable.

In view of the dominant role of $\pi^+\pi^-$ for the evaluation of 
$a_\mu^\mathrm{had,LO}$ we will from now on concentrate on this 
specific final state. In Section~5 we will argue that the  
measurement of $R(\pi^+\pi^-(\gamma))$ is
indeed possible with the technique of the radiative return. Before
entering this discussion, a precise definition of the objects of
interest is required. We will rely on the smallness of the
fine structure constant and only consider NLO amplitudes and rates. 
In leading order the contribution to $R$ and
$a_\mu^\mathrm{had,LO}$ is given by the square of the pion form factor. 
In most experiments the issue of contributions from events with
additional photon emission from the hadronic system and the influence
of these photons on cuts has simply been ignored (for a first step
towards inclusion and control of these effects 
from the experimental side, see~\cite{CMD2}; for related 
theoretical discussions, see~\cite{Melnikov:2001uw,Gluza:rad}).
These effects indeed are only relevant in next-to-leading order.

To proceed to ${\cal O}(\alpha)$ for point-like particles such as muons or
electrons, we simply evaluate corrections to the vertex function from
virtual photon exchange, which must be combined with real
radiation to arrive at an infrared-finite result. For pions we must
follow a different strategy. The evaluation of virtual corrections
involves the full hadronic dynamics and seems to be difficult, if not
impossible, but is also unnecessary. Instead, we separate the inclusive rate
$R(\pi^+\pi^-(\gamma))$ into a part containing the virtual plus real photonic
corrections $R^{\mathrm{V+R}}(\pi^+\pi^-(\gamma), E^{\mathrm{cut}})$,
with energies up to a cutoff \({E}^{\mathrm{cut}}\) chosen
such that the point-like pion model describes real emission, say up to 50 
or even 100~MeV, and a remainder
$R^{\mathrm{H}}(\pi^+\pi^-\gamma,E^{\mathrm{cut}})$
with hard photons only. The separation between the two configurations will
depend on the cutoff $E^\mathrm{cut}$, which has to be chosen small for
this dependence to be given by the familiar result for point-like pions. 

\begin{figure}[ht]
\epsfig{file=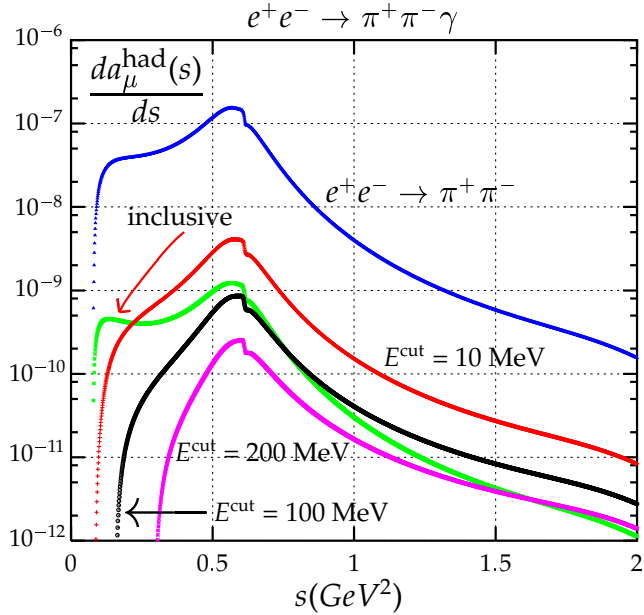,width=8.5cm} 
\caption{Differential contribution to $a_\mu^\mathrm{had,\gamma}$ 
from $\pi^+\pi^-\gamma$ intermediate states for different cutoff values 
compared with the complete contribution (virtual plus real corrections, 
labelled `inclusive') evaluated in sQED (FSR), as well as 
with the contribution from the $\pi^+\pi^-$ intermediate state.}
\label{fig6}
\end{figure}

\begin{figure}[ht]
\epsfig{file=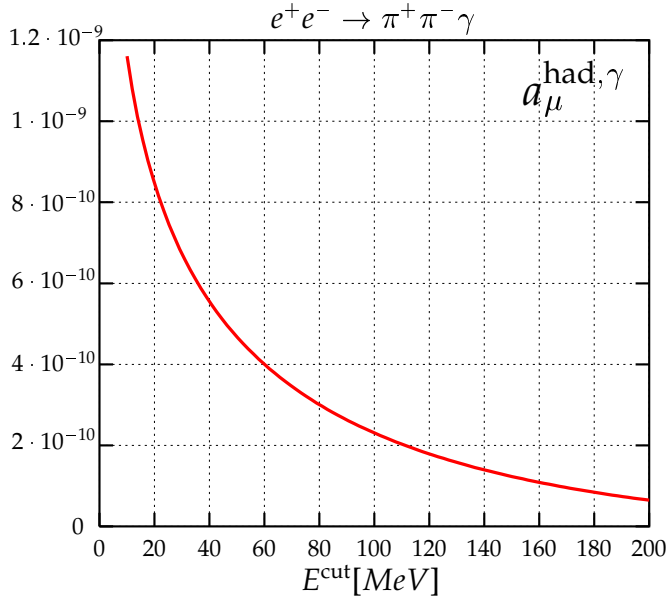,width=8.8cm}
\caption{Integrated contribution to $a_\mu^\mathrm{had,\gamma}$  
as a function of the cutoff $E^\mathrm{cut}$.}
\label{fig7}
\end{figure}

The strategy for the exclusive measurement now proceeds as
follows:

The contribution from $R^{\mathrm{V+R}}(\pi^+\pi^-(\gamma), E^{\mathrm{cut}})$
as a function of $E^\mathrm{cut}$ can be measured by analysing final states 
where hard-photon radiation is suppressed by suitable chosen cuts on energy 
and collinearity of the pions. The $E^\mathrm{cut}$ dependence of 
$R^{\mathrm{H}}(\pi^+\pi^-\gamma,E^{\mathrm{cut}})$ can be measured
by collecting $\pi^+\pi^-\gamma$ final states (and correcting of
course for ISR) as discussed in Section~\ref{sec:return}. 
Alternatively we may correct for (unmeasured) hard photon events by 
employing a model like sQED, which has to be checked
experimentally --- at least for a few selected kinematic
configurations.

Let us now estimate the contributions from hard photon radiation to
$a_\mu$. In Fig.~\ref{fig6} we display the integrand
\begin{equation}
\frac{d}{ds} a^\mathrm{had,\gamma}_\mu(E^\mathrm{cut}) = 
\frac{\alpha^2}{3\pi^2 s} K(s) \ R^{\mathrm{H}}(s,E^\mathrm{cut})~,
\end{equation}
for $E^\mathrm{cut} =10$, $100$ and $200$~MeV as a function of $\sqrt{s}=
m(\pi^+\pi^-\gamma)$ between the threshold and $2$~GeV. The result is
compared with the complete sQED contribution, as derived from point-like
pions, and the lowest order contribution from $\pi^+\pi^-$. 
The integrated result is displayed in Fig.~\ref{fig7}.

Using sQED as a model, one finds that contributions from 
hard photon radiation, with a cut at $100$~MeV, are still small with respect 
to the present uncertainty $\delta a_\mu^\mathrm{had,LO} = 7 \times 10^{-10}$
\cite{Davier:2002dy,HMNT02}. Let us emphasize that cuts on the photon
energy around 50~MeV or below might well lead to important shifts 
in \(a_\mu\). Of course we have assumed that
hard radiation is not grossly underestimated by sQED, an assumption to
be tested by experiment. As we will demonstrate in the next section, such
tests are indeed feasible in experiments based on the radiative return.

\section{FSR at next-to-leading order and the event generator PHO\-KHA\-RA}
\label{sec:fsrnlo}

As discussed in Section~\ref{sec:return}, the radiative return allows us 
to exploit the enormous luminosity of $\Phi$- and $B$-meson factories for the
measurement of the hadronic cross section over a wide range of energies.
On the basis of the leading order treatment, we have demonstrated that
the analysis of final states with one photon allows us
to determine the pion form factor for arbitrary $Q^2$ and the cross
section for $\pi^+\pi^-$ plus a hard photon from FSR at $\sqrt{s}$.

Intuitively it should also be possible to exploit the radiative return
for the extraction of FSR for arbitrary invariant mass $\sqrt{s'}$
of the $\pi^+\pi^-\gamma$ system through the reaction
\begin{equation}
e^+e^-\to \gamma\gamma^*(\to \pi^+\pi^-\gamma) \ .
\label{eq:twostep}
\end{equation}
The implementation of this two-step process into PHO\-KHA\-RA and the
question of how to separate the corresponding amplitude from double
photon emission from the initial or final state are the main topics of
this section.

\begin{figure}[ht]
\epsfig{file=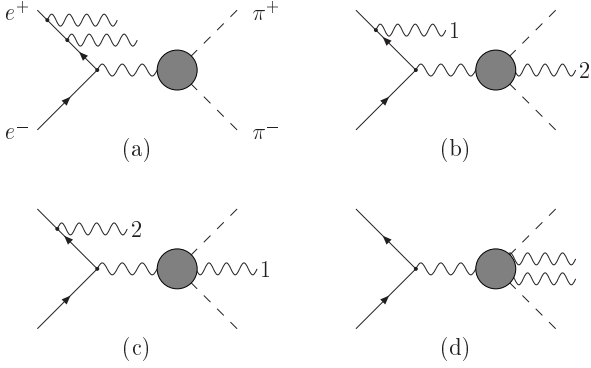,width=8.5cm} 
\caption{Typical amplitudes contributing to the reaction
$e^+e^-\to\pi^+\pi^-\gamma\gamma$. For two photons emitted either from
the electron/positron or the hadronic system. Only one representative is
displayed.}
\label{fig8}
\end{figure}

\begin{figure}[ht]
\epsfig{file=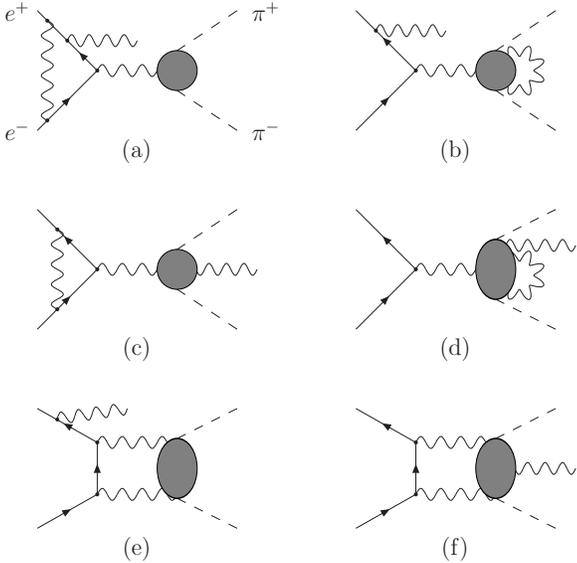,width=8.5cm} 
\caption{Typical amplitudes describing virtual corrections to the reaction
$e^+e^-\to\pi^+\pi^-\gamma$. Permutations are omitted.}
\label{fig9}
\end{figure}

\begin{figure}[ht]
\begin{center}
\hspace{-.3cm}\epsfig{file=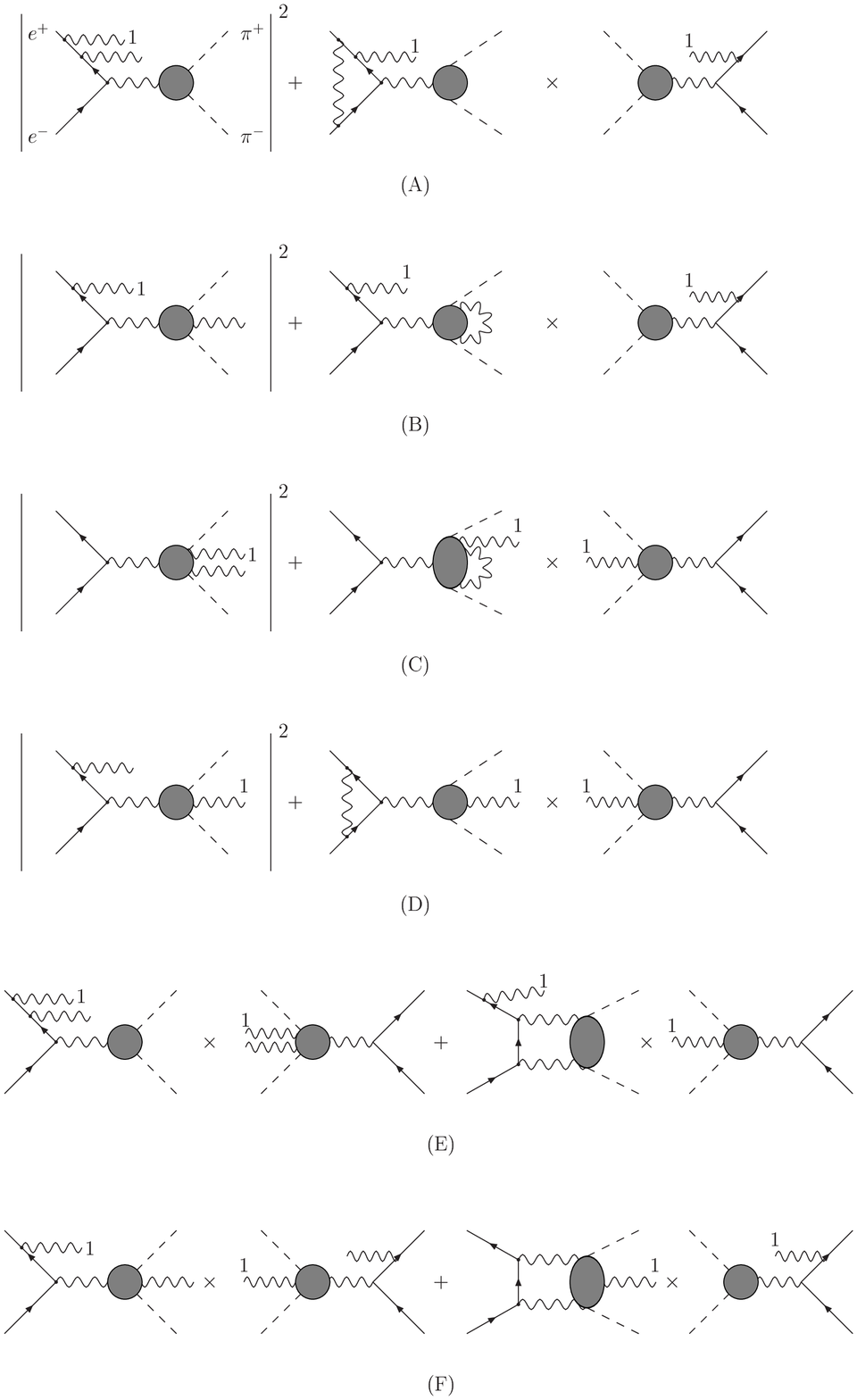,width=9cm} 
\caption{Charge-even infrared-finite combinations of amplitudes.
The photon labelled `1' is hard. Diagrams written in reverse order are to 
be understood as complex conjugate.}
\label{fig10}
\end{center}
\end{figure}

Let us first describe the main physics ingredients and assumptions of
the new version of PHO\-KHA\-RA (PHO\-KHA\-RA 3.0).
The complete set of diagrams relevant to the reaction
\begin{equation}
e^+e^-\to \pi^+\pi^-\gamma\gamma~,
\end{equation}
is displayed in Fig.~\ref{fig8}. We require at least one photon 
(with four momentum  $k_1$) to be
hard, i.e. with energy $E_1> E_\gamma^\mathrm{min}$, where  
$E_\gamma^\mathrm{min}$ is typically around $5$ to $10$~MeV for 
\DAF \ energies and significantly larger (several GeV) 
for $B$-meson factories. We only display typical diagrams and omit
permutations. However, we explicitly distinguish the two cases, where the
hard photon is emitted either from the electron--positron line or from
the hadronic system. Diagrams relevant to virtual corrections to the 
reaction with one photon,
\begin{equation}
e^+e^-\to \pi^+\pi^-\gamma~,
\end{equation}
are shown in Fig.~\ref{fig9}; they contribute through their interference 
with the Born amplitude (Fig.~\ref{fig1}).
At this point no assumption about the
form of the $\gamma^*\to \pi^+\pi^-\gamma$ coupling or the structure
of the virtual corrections at the hadronic vertex is imposed.

Just as in leading order, charge-symmetric and asymmetric terms are
present in the differential distribution. The latter are particularly 
sensitive to the FSR amplitude and may serve for tests of the
FSR model or even for an independent determination of this amplitude
for arbitrary $s'$. In the present upgraded version of the program,
only charge symmetric terms will be included, which are sufficient for
the present purpose.

In Fig.~\ref{fig10} those combinations of (gauge-invariant sets of) amplitudes 
involving real and virtual radiation are displayed, which lead to 
infrared-finite charge-even combinations. The first (A)
corresponds to the radiative corrections to the leptonic tensor, and has
been derived in \cite{Rodrigo:2001jr,Kuhn:2002xg}. It is the crucial element
of PHO\-KHA\-RA (versions 1.0 and 2.0) and can be used for the simulation of
arbitrary hadronic final states. The second combination (B) corresponds to
the two-step process described in~\Eq{eq:twostep} above and is 
implemented in the present, newest version of PHO\-KHA\-RA (version 3.0). 
Before discussing the actual implementation, let us discuss
the remaining terms. The next combinations, denoted by (C) and (D),
describe radiative corrections to pure FSR, with the leading process
depicted in Fig.~\ref{fig1}b. The hard photon originates from the
hadronic system, and the radiative corrections (virtual, soft, and hard
radiation) are either attached to the hadronic (C) or the leptonic (D)
system. As discussed in Section~\ref{sec:return}, 
the leading order process can be controlled at \DAF \
through suitable cuts and FSR can be reduced to a level of less than 1\%.  
At the energy of $10.52$~GeV, i.e. at $B$-meson factories, 
it is entirely irrelevant. Therefore we do not include these radiative
corrections to an already small contribution in the program.

A similar argument applies to the remaining terms, denoted by (E) and (F). 
These describe radiative corrections (real and virtual photons) to the
interference between hard ISR and hard FSR. They will be neglected on
the basis of the same arguments as before.

In the following, we denote by ISRNLO the contributions to the 
cross section coming from ISR only calculated at NLO, i.e. those 
from Figs.~\ref{fig1}a and~\ref{fig10}A. IFSLO includes 
also the contribution from LO FSR (diagrams in Figs.~\ref{fig1}a, 
\ref{fig1}b and~\ref{fig10}A).
Finally, IFSNLO includes on top of that initial plus final state emission at NLO
(diagrams in Figs.~\ref{fig1}a, \ref{fig1}b, \ref{fig10}A and~\ref{fig10}B).
Let us emphasize that these are gauge-invariant subsets throughout. 

Let us now describe the implementation of the two-step 
process~\Eq{eq:twostep}, denoted IFSNLO, in detail.
Since we are only interested in final states with masses of
the $\pi^+\pi^-\gamma$ system below $\approx$ 1 GeV, we expect sQED to
provide a good description of the $\gamma^*\to \pi^+\pi^-\gamma$ amplitude.
The virtual corrections  are similarly modelled by sQED. They are combined 
with soft radiation, with $E_2<E_2^\mathrm{cut}$, 
to arrive at an infrared-finite result. The effective form factor is, 
therefore, introduced through the substitution
\begin{eqnarray}
|F^0(Q^2)|^2 &\to& |F(Q^2,E_2^\mathrm{cut})|^2 =  \non \\ & &
|F^0(Q^2)|^2 \left(1+\frac{\alpha}{\pi} \eta^\mathrm{V+S}
(Q^2,E_2^\mathrm{cut}) \right) \ , \ 
\label{newff}
\end{eqnarray}
with $\eta^\mathrm{V+S}(Q^2,E_2^\mathrm{cut})$ given below. For the 
generation of real radiation the cutoff energy obviously refers to the
$\pi^+\pi^-\gamma$ centre-of-mass frame.

It should be emphasized that only the combination $|F(Q^2,E_2^\mathrm{cut})|^2$
is physically observable and relevant to the analysis of $a_\mu$ and
$\alpha(m_Z)$, of course after adding real radiation, which can either
be measured, or for photon energies up to \(\cal O\)(100~MeV), 
modelled by sQED through the reaction
\bea
&&\kern-20pt 
 e^+(p_1,\lambda_{e^+}) \ + \ e^-(p_2,\lambda_{e^-}) \to \nonumber \\
&& \kern10pt
 \pi^+(q_1) \ + \ \pi^-(q_2) \ + \ \gamma(k_1,\lambda_1)
  \ + \ \gamma(k_2,\lambda_2)~. \nonumber \\
\label{eeppgg}
\eea

The helicity amplitudes responsible for real radiation 
(Fig.~\ref{fig8}b and c) are given by the sum of 12 Feynman amplitudes. 
Adopting a notation similar to~\cite{Rodrigo:2001kf}, they read 
\begin{align}
 &M^\mathrm{H}_\mathrm{IFSNLO}(\lambda_{e^+},\lambda_{e^-},\lambda_1,\lambda_2) =
 -\frac{(4\pi\alpha)^2}{\hat Q^{2}} \biggl\{ \non \\ &
 v_I^{\dagger}(p_1,\lambda_{e^+}) \
 A \ u_I(p_2,\lambda_{e^-})+v_{II}^{\dagger}(p_1,\lambda_{e^+}) \
 B \ u_{II}(p_2,\lambda_{e^-}) \biggl\} \non \\ &
  + \ \ \  \bigl( k_1 \leftrightarrow k_2 \bigr)
 \ \ \ ,
\end{align}
where the index H indicates that both photons are hard, and 
\begin{align}
 A &= 
\frac{\left(\varepsilon^*(k_1,\lambda_1)^- k_1^+ 
 - 2\varepsilon^*(k_1,\lambda_1)\cdot p_1\right)J^{-}}
 {2 k_1 \cdot p_1}\non \\ &
 +  \frac{J^{-}\left(2\varepsilon^*(k_1,\lambda_1)\cdot p_2 
 -  k_1^+ \varepsilon^*(k_1,\lambda_1)^-\right)}{2 k_1 \cdot p_2}~,
\end{align}
\begin{align}
 B &= 
\frac{\left(\varepsilon^*(k_1,\lambda_1)^+ k_1^- 
 - 2\varepsilon^*(k_1,\lambda_1)\cdot p_1\right)J^{+}}
 {2 k_1 \cdot p_1}\non \\ &
 +  \frac{J^{+}\left(2\varepsilon^*(k_1,\lambda_1)\cdot p_2 
 -  k_1^- \varepsilon^*(k_1,\lambda_1)^+\right)}{2 k_1 \cdot p_2}~,
\end{align}
with 
\begin{align}
\hat Q = p_1 + p_2 - k_1 = q_1 + q_2 + k_2~, \qquad s' = \hat Q^2~,
 \end{align}
and \(J^\mu\) being the current describing the \(\pi^+ \pi^-\gamma\) 
final state 
\begin{align}
J^{\mu} = i F_{2\pi}(s') \ D^{\mu}~,
\end{align}
where
\begin{align}
&D^{\mu} = \left(q_1+k_2-q_2\right)^{\mu} 
 \frac{q_1\cdot\varepsilon^*(k_2,\lambda_2)}{q_1\cdot k_2}\non \\ &
 +\left(q_2+k_2-q_1\right)^{\mu} 
 \frac{q_2\cdot\varepsilon^*(k_2,\lambda_2)}{q_2\cdot k_2}
 -2\varepsilon^{*\mu}(k_2,\lambda_2) \ ,
\end{align}
and \(k^\pm,J^\pm\) are 2$\times$2 matrices, which can be found 
in~\cite{Rodrigo:2001kf}. We do not include contributions from the 
radiative decay $\phi \to \pi^+\pi^-\gamma$. Their effect is small, 
can be well controlled and was discussed in \cite{Melnikov:2000gs}.

In the present version of the program, one of the photons is assumed
to be visible, at least in principle, and only the photon emitted from final states
is allowed to be soft (Fig.~\ref{fig10}B). For photon energies
$w\sqrt{s}<E_{1\gamma}<E_{c}^{v} $ (or $w\sqrt{s}<E_{2\gamma}<E_{c}^{v}$)
the amplitude used consists of 6 diagrams only. $E_{c}^{v}$ is the threshold 
above which we can `observe' the photon --- typically $10$ MeV for \DAF \
and 100 MeV for $B$-factories. If both photons are hard,
$E_{1\gamma},~E_{2\gamma}>E_{c}^{v}$ the sum of 12 diagrams is used. 

The corresponding virtual plus soft photon corrections can be written as
\bea
d\sigma^\mathrm{V+S}_\mathrm{IFSNLO} = \frac{\alpha}{\pi} \ 
   \eta^\mathrm{V+S}(s',E_2^\mathrm{cut}) \ 
   d\sigma^{(0)}_{\mathrm{ISR}}(s')~,
\eea
where $d\sigma^{(0)}_{\mathrm{ISR}}$ is the leading order 
\(e^+e^- \to \pi^+\pi^-\gamma\) cross section, with the photon 
emitted off the initial leptons only, and
\begin{align}
& \eta^\mathrm{V+S}(s',E_2^\mathrm{cut})  = 
- 2 \biggl[ \frac{1+\beta^2}
{2\beta} \log(t) + 1 \biggr] 
\log(2w)  \nonumber \\
&- \frac{2+\beta^2}{\beta} \log(t) - 2 
+ \log\biggl(\frac{1-\beta^2}{4}\biggr) 
\nonumber \\ &
- \frac{1+\beta^2}{2\beta} \biggl\{ -\log(t) 
\log\biggl(\frac{\beta(1+\beta)}{2}\biggr)
\nonumber \\ &
+ \log\biggl(\frac{1+\beta}{2\beta}\biggr) \log\biggl(\frac{1-\beta}
{2\beta}\biggr) 
\nonumber \\ &
+ 2\Li_2 \biggl( \frac{2\beta}{1+\beta} \biggr)
+ 2\Li_2 \biggl( -\frac{1-\beta}{2\beta} \biggr)
- \frac{2}{3}\pi^2 \biggr\}~, \nonumber \\ 
\label{etavs}
\end{align}
where $w = E_{2}^{\mathrm{cut}}/\sqrt{s'}$, and 
\bea
\beta = \sqrt{1 - 4m_{\pi}^{2}/s'}~, \qquad t = \frac{1-\beta}{1+\beta}~.
\label{beta}
\eea

The function $\eta^\mathrm{V+S}(s',E_2^\mathrm{cut})$ is of course the familiar
correction factor derived in~\cite{Schwinger:ix} for the reaction 
$e^+e^- \to \pi^+ \pi^- \gamma$ in the framework of sQED 
(see also \cite{Gluza:eta}).
For the present case, $s'$ corresponds to the squared mass of the 
$\pi^+ \pi^- \gamma$ system, and for virtual and soft photon 
emission, $s' \simeq Q^2$, the soft photon cutoff is defined 
in the $\pi^+ \pi^- \gamma$ rest frame.

Correspondingly the cross section for the reaction 
\begin{equation}
e^+ e^- \to \gamma(k_1) \gamma^* (\to \pi^+ \pi^- \gamma(k_2))~,
\end{equation}
after integration over the angles and energy
(from $E_2^\mathrm{cut}$ to the kinematical limit) of $\gamma(k_2)$ ,
is given by 
\bea
d\sigma^\mathrm{H}_\mathrm{IFSNLO} = \frac{\alpha}{\pi} \ 
   \eta^\mathrm{H}(s',E_2^\mathrm{cut}) \ 
   d\sigma^{(0)}_{\mathrm{ISR}}(s')~,
\eea
with
\begin{align}
& \eta^\mathrm{H}(s',E_2^\mathrm{cut}) =
- \frac{1+\beta^2}{\beta}\biggl[\Li_{2}\biggl( 1-\frac{t_{m}}{t} \biggr)
 - \Li_{2}(t_{m}t) \nonumber \\&
+ \zeta(2) + \frac{\log^{2}(t)}{2}\biggr] 
- \log(t_{m}-t)\biggl[\frac{1+\beta^2}{\beta}\log \biggl(\frac{t_{m}}{t}
 \biggr)  \nonumber \\&
-2 \biggr] - \frac{(1-\beta^2)\beta_{m}}{2\beta^3(1-\beta_{m}^{2})}
\biggl[ \frac{2(1-\beta^2)}{1-\beta_{m}^2} - 7\beta^2 - 5\biggr] \nonumber \\&
+ \frac{\log(t_m)}{\beta}\biggl\{ -2 + \frac{1-\beta^2}{4}\biggl[
1 + \frac{8}{1-\beta_{m}^2} + \frac{3}{\beta^2}\biggr] \nonumber \\&
+ (1+\beta^2)\biggl[  \log\biggl( \frac{4(\beta^2 - \beta_{m}^2)}
{(1+\beta)^2(1-\beta_{m}^2)}  \biggr) 
+ \frac{\log(t_m)}{2}  \biggr]   \biggr\} \nonumber \\&  - 2 \log(1-t_mt) ~,
\label{etah}
\end{align}
where $t$ and $\beta$ are defined in \Eq{beta}, and
\bea
\beta_m \equiv \sqrt{1-\frac{4m^2_\pi}{Q_{m}^{2}}}~, \qquad 
t_{m} = \frac{1-\beta_m}{1+\beta_m}~. \nonumber  
\eea
Here \(Q_{m}^{2}\) is the maximum value of \(Q^2\) 
\begin{equation}
Q_{m}^{2} = s' - 2 E_2^{\mathrm{cut}} \sqrt{s'}~,  \nonumber  
\end{equation}
and $E_2^\mathrm{cut}$ is again defined in the $\gamma^*$ rest frame. 
This formula will be useful for tests of the Monte Carlo event generator
discussed below, and $E_2^\mathrm{cut}$ is still arbitrary at 
this point. The same formula can also be used to test FSR at LO 
with $s'\to s$ and $E_2^\mathrm{cut}$ set now in the $e^+e^-$ cms frame.

For $E_2^\mathrm{cut}$ small, $w = E_2^\mathrm{cut}/\sqrt{s'} \ll 1$,
the function $\eta^\mathrm{H}$ reduces to
\begin{align}
& \kern-5pt \eta^\mathrm{H}(s',E_2^\mathrm{cut}) \simeq 
 \log(2w) 
\biggl[2+\frac{1+\beta^{2}}{\beta}\log(t)\biggr] \label{etahsoft}\\ 
& \kern-5pt 
 - \frac{(1+\beta^{2})}{\beta}
 \biggl[ \log(t^2) \log(1-t)
   + \Li_{2}(1-t^{2}) \biggr]
 \nonumber \\ &\kern-5pt
+ \frac{(1-\beta^2)(3+\beta^2)}{4\beta^3}\log(t)   
 + 2 \log\biggl(\frac{1-\beta^2}{4\beta^2}\biggr) 
 + \frac{3}{2\beta^{2}} +\frac{7}{2} ~. 
\nonumber 
\end{align}
Adding virtual, soft (\Eq{etavs}) and hard (\Eq{etahsoft}) corrections, 
the familiar correction factor~\cite{Schwinger:ix,Drees:1990te,Melnikov:2001uw}
\begin{align}
& \eta(s') = \frac{1+\beta^2}{\beta} \biggl\{ 4\Li_{2}
\biggl( \frac{1-\beta}{1+\beta} \biggr) 
+ 2\Li_{2} \biggl( - \frac{1-\beta}{1+\beta}\biggr) \nonumber \\
& - 3\log\biggl( \frac{2}{1+\beta}\biggr)
 \log\biggl( \frac{1+\beta}{1-\beta}\biggr)
- 2\log(\beta)\log\biggl( \frac{1+\beta}{1-\beta} \biggr)\biggr\} \nonumber \\
& + \frac{1}{\beta^3} \biggl[ \frac{5}{4} (1+\beta^2)^2 - 2\biggr]
\log\biggl(\frac{1+\beta}{1-\beta}\biggr) 
+ 3\log\biggl(\frac{1-\beta^2}{4}\biggr) 
\nonumber \\
& - 4\log(\beta) + \frac{3~(1+\beta^2)}{2\beta^2}~,
\label{etatot}
\end{align}
is recovered.

Using Eqs.~(\ref{etavs}) and (\ref{etah}) the implementation of FSR in 
combination with ISR is straightforward. To match hard, soft and 
virtual radiation smoothly, the energy cutoff (\(E_2^\mathrm{cut}\))
has to be transformed from the rest frame of the 
$\pi^+ \pi^- \gamma$ (emitted from the final state) system  to 
the laboratory frame (\(e^+e^-\) cms frame) (\(E_{\gamma}^\mathrm{min}\)).
In fact it is necessary to recalculate the soft photon contribution,
as the cut on \(Q^2\) depends in the latter case on the angle between
the two emitted photons and now
\begin{align}
& \eta^\mathrm{V+S}(s,s',E_{\gamma}^\mathrm{min})  = 
- 2 \biggl[ \frac{1+\beta^2}
{2\beta} \log(t) + 1 \biggr] 
\nonumber \\ & \times
\biggl[ \log(2w) + 1 + \frac{s'}{s'-s} 
\log\biggl(\frac{s}{s'} \biggr) \biggr] \nonumber \\
&- \frac{2+\beta^2}{\beta} \log(t) - 2 
+ \log\biggl(\frac{1-\beta^2}{4}\biggr) 
\nonumber \\ &
- \frac{1+\beta^2}{2\beta} \biggl\{ - \log(t) 
 \log\biggl(\frac{\beta(1+\beta)}{2}\biggr) 
\nonumber \\ &
+ \log\biggl(\frac{1+\beta}{2\beta}\biggr) 
\log\biggl(\frac{1-\beta}{2\beta}\biggr) 
\nonumber \\ &
+ 2\Li_2 \biggl( \frac{2\beta}{1+\beta} \biggr)
+ 2\Li_2 \biggl( -\frac{1-\beta}{2\beta} \biggr)
- \frac{2}{3}\pi^2 \biggr\}~, 
\label{etavsp}
\end{align}
where $\beta$ is defined in \Eq{beta} and 
$w = E_{\gamma}^\mathrm{min}/\sqrt{s}$ . 

\begin{table}[t]
\caption{Total cross section (nb) for the process
$e^+ e^- \rightarrow \pi^+ \pi^- \gamma(\gamma)$ for different values
of the soft photon cutoff. One of the photons is required to have an 
energy larger than 10 MeV (100 MeV) for $\sqrt{s} = 1.02$~GeV (10.52~GeV). 
No further cuts applied in columns 2 and 3. In column 4,
the pion angles are restricted to $40^\circ < \theta_{\pi^\pm} < 140^\circ$
and the photon(s) angles to $\theta_{\gamma} < 15^\circ$
or $\theta_{\gamma} > 165^\circ$.}
\label{tab:epstest}
\begin{center}
\begin{tabular}{cccc}
\(w\) & \(\sqrt{s}=\)1.02~GeV  & 10.52~GeV & 1.02~GeV \\  \hline 
$10^{-3}$ & 40.992 (5)  & 0.1606 (1) & 20.988 (1) \\
$10^{-4}$ & 41.013 (6)  & 0.1607 (2) & 20.993 (1) \\
$10^{-5}$ & 41.018 (7)  & 0.1607 (2) & 20.995 (2) \\ \hline 
\end{tabular}
\end{center}
\end{table}

\begin{figure}[ht]
\begin{center}
\epsfig{file=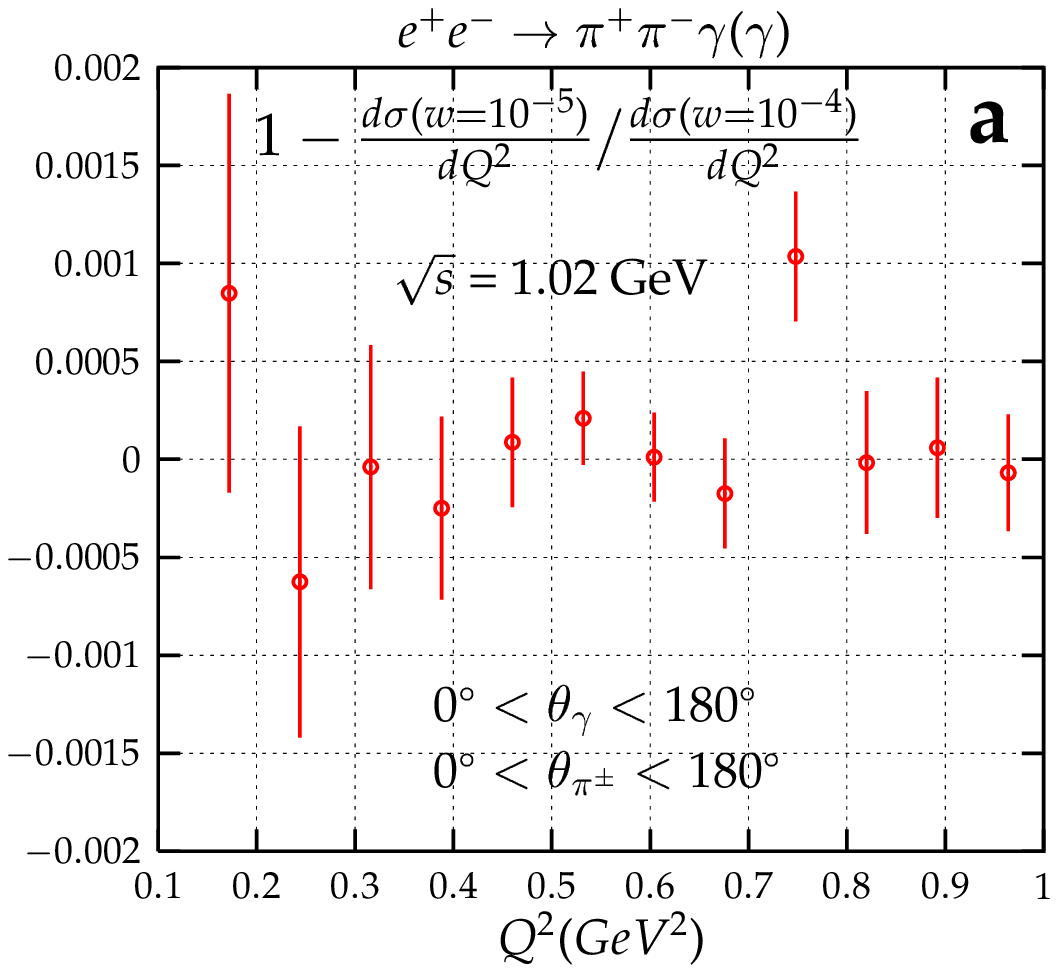,height=6cm,width=8.cm}
\epsfig{file=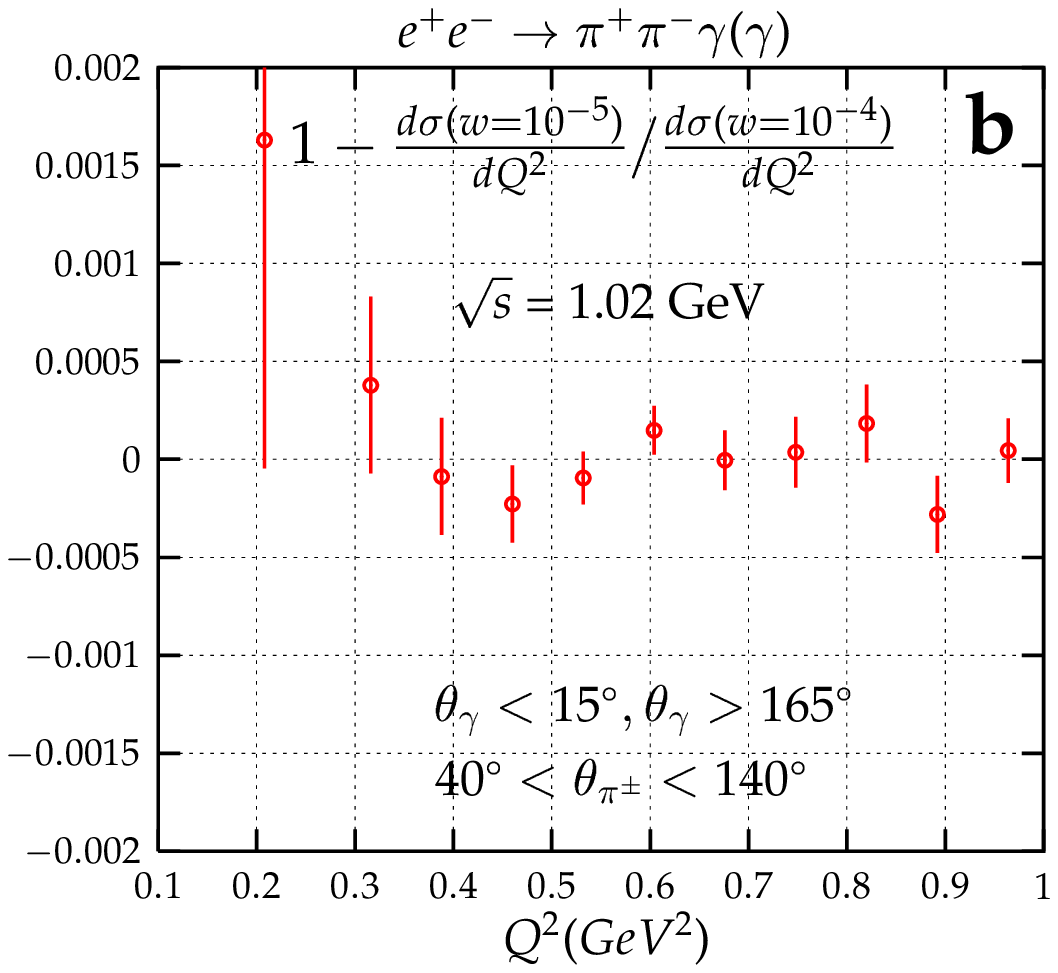,height=6cm,width=8.cm}
\epsfig{file=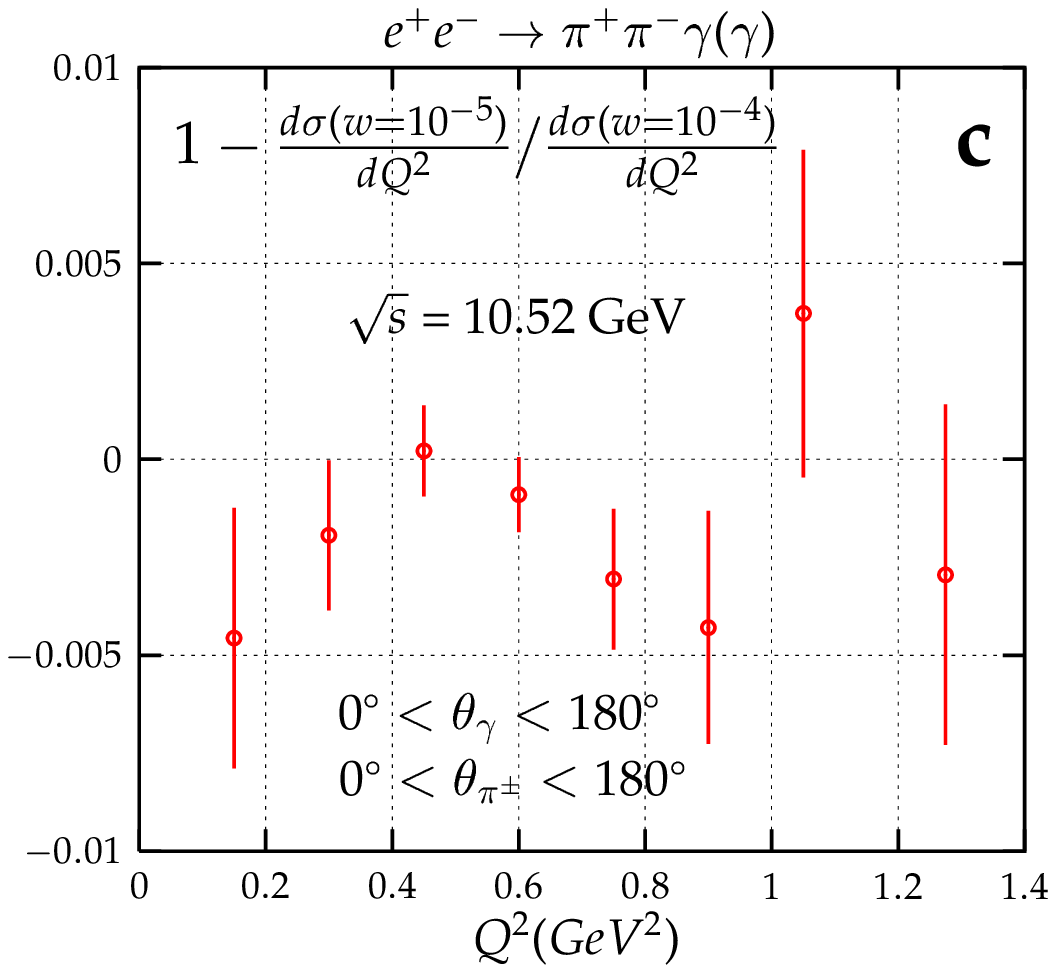,height=6cm,width=8.cm}
\caption{Comparison of the $Q^2$ distribution for two
values of the soft photon cutoff ($w = 10^{-4}$ vs. $10^{-5}$)
for $\sqrt{s}=1.02$~GeV (a,b) and $\sqrt{s}=10.52$~GeV (c). 
One of the photons was required to have energy $>$ 10~MeV 
(for $\sqrt{s}=1.02$~GeV) and $>$ 100~MeV (for $\sqrt{s}=10.52$~GeV). 
Angular cuts were applied only on plot (b), where pion polar angles are 
restricted to $40^\circ < \theta_{\pi^\pm} < 140^\circ$ and
photon polar angles to $\theta_{\gamma} < 15^\circ$ and $\theta_{\gamma} > 165^\circ$.}
\label{fig:cut_10.52}
\vspace{-1cm}
\end{center}
\end{figure} 

\begin{figure}[ht]
\begin{center}
\epsfig{file=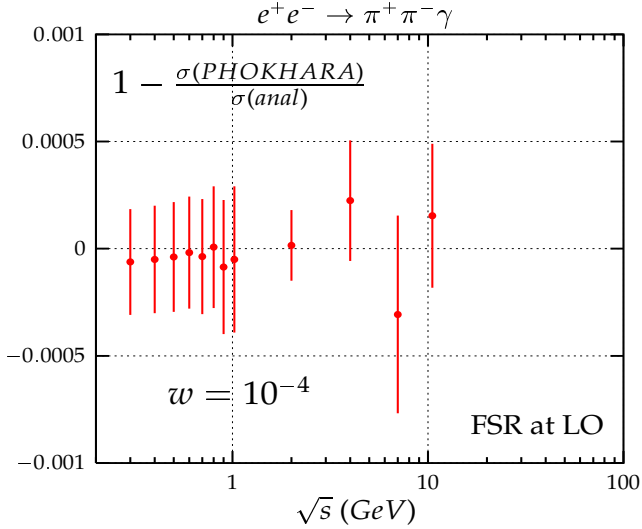,height=7cm,width=8.5cm}
\caption{Comparison between the LO FSR cross section calculated analytically 
(\Eq{sig_1ph}) and calculated by PHO\-KHA\-RA for fixed
value of $w=E_{\gamma}^\mathrm{min}/\sqrt{s}$.}
\label{fig:eg_min_c}
\end{center}
\end{figure}

\begin{figure}[ht]
\begin{center}
\epsfig{file=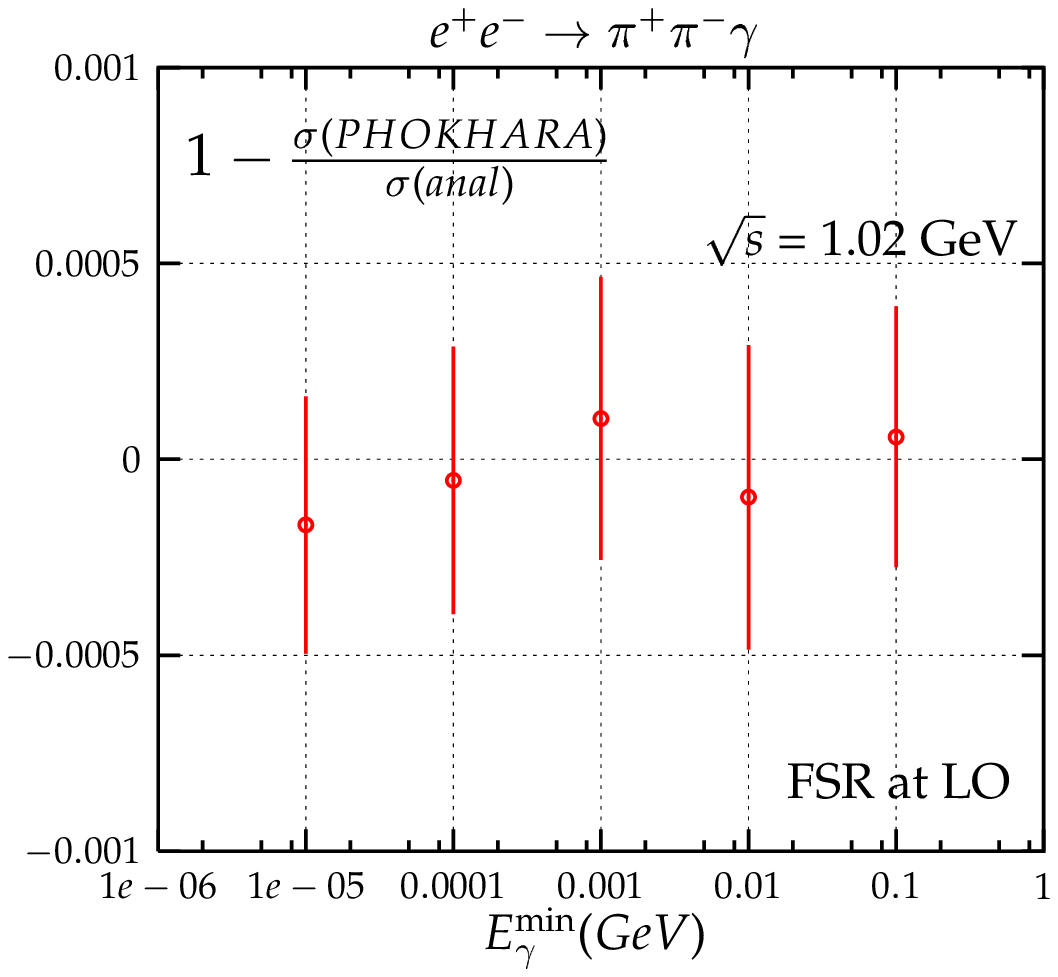,height=7cm,width=8.5cm}
\epsfig{file=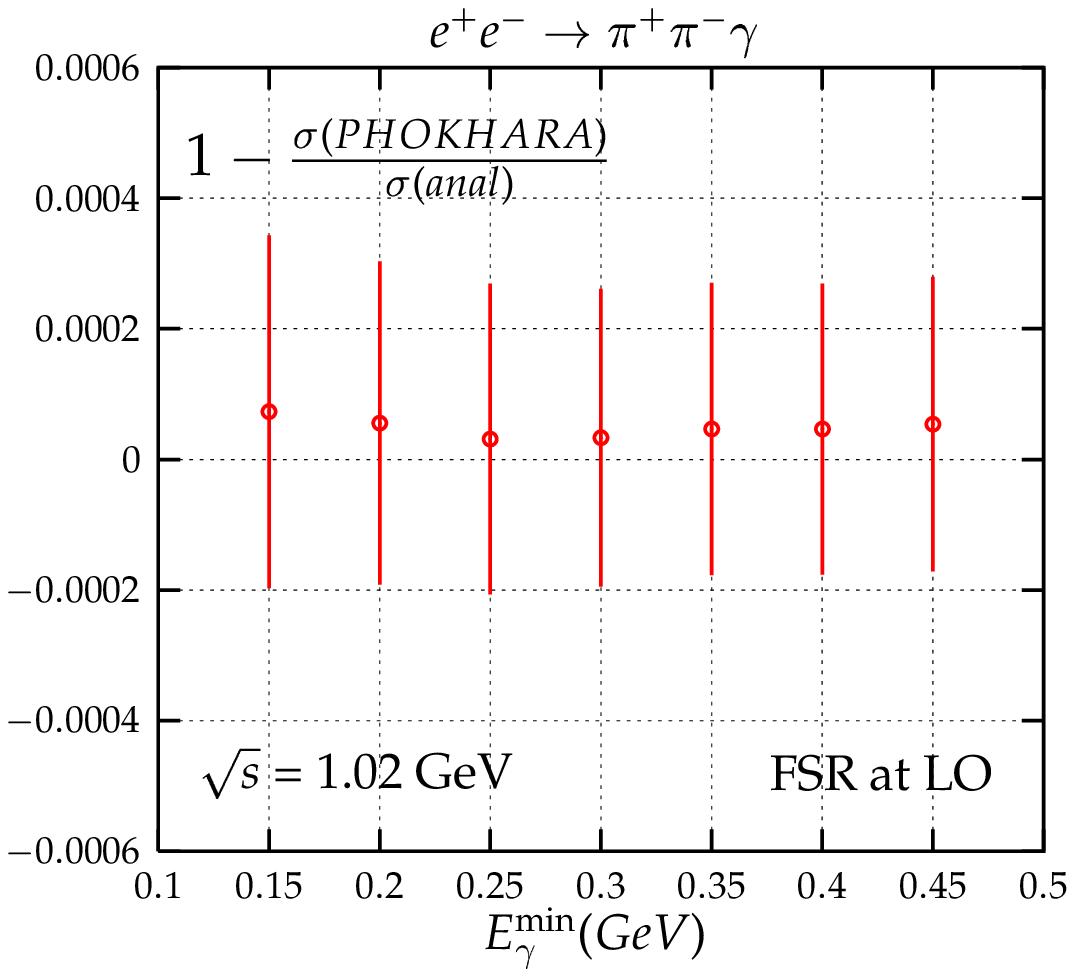,height=7cm,width=8.5cm}
\caption{Comparison between the LO FSR cross section calculated analytically 
(\Eq{sig_1ph}) and calculated by PHO\-KHA\-RA for a fixed
value of \( \sqrt{s} = 1.02 \)~GeV.}
\label{fig:eg_min_1.02}
\end{center}
\end{figure}

A number of tests were performed to ensure the technical 
precision of the new version of PHO\-KHA\-RA. The square of the matrix 
element summed over polarizations of the final particles
and averaged over polarizations of the initial particles 
was calculated with FORM~\cite{FORM} using the standard trace method.
External gauge invariance was checked analytically when using the trace
method and numerically for the amplitude calculated with the helicity
am\-pli\-tu\-de me\-thod. The two results for the square of the matrix element
summed over polarizations, were compared numerically. The code based
on the result from the trace method was written in quadrupole precision
to reduce cancellations. The code based on the result obtained with the helicity 
amplitude method, uses double-precision for real and complex numbers and is now 
incorporated in the code of PHO\-KHA\-RA 3.0. Agreement of 13 significant 
digits (or better) was found between both codes.
The sensitivity of the integrated cross section --- with and without angular cuts ---
to the choice of the cutoff $w$ can be deduced from Table~\ref{tab:epstest}
and Fig.~\ref{fig:cut_10.52}.
For simplicity the same separation parameter $w$ was chosen for ISR and FSR 
corrections. Choosing $w=10^{-4}$ or less, the result becomes independent of $w$. 
The tests prove that the analytical formula describing soft photon emission as well as
the Monte Carlo integration in the soft photon region are well implemented
in the program. Having analytical expressions for 
\(\eta^\mathrm{H}(s,E_{\gamma}^\mathrm{min})\) (\Eq{etah})
we can test also the implementation of hard photon emission
from the final state. The results of the tests are collected 
in Figs.~\ref{fig:eg_min_c} and~\ref{fig:eg_min_1.02}, where 
FSR at LO obtained from PHOKHARA is compared with the analytical
result for the corresponding cross section:
\begin{align}
 \sigma_\mathrm{FSR}(s)  &= \frac{\alpha}{\pi} \ 
 \eta^\mathrm{H}(s,E_{\gamma}^{\mathrm{min}}) \ \sigma_0(s)& \nonumber \\
   &= \frac{\alpha}{\pi}
 \ \eta^\mathrm{H}(s,E_{\gamma}^{\mathrm{min}})  \ \frac{\pi}{3}
 \frac{\alpha^2 \beta^3(s)}{s}\mid F_{2\pi}(s)\mid^{2}~,
\label{sig_1ph}
\end{align}
for a fixed value of the photon energy cutoff and different values of the 
\(e^+e^-\) cms energy (Fig.~\ref{fig:eg_min_c}) or fixed value of the \(e^+e^-\) 
cms energy and various values of the photon energy cutoff (Fig.~\ref{fig:eg_min_1.02}).
As one can see the technical precision of that part of PHO\-KHA\-RA is much better 
than 1 per mille. The IFSNLO part was also tested against the analytical result
\begin{align}
 \eta(s')   = \frac{d\sigma^\mathrm{V+S+H}_\mathrm{IFSNLO}(s,s')/ds'} {\frac{\alpha}{\pi} \ 
    d\sigma^{(0)}_{\mathrm{ISR}}(s')/ds'} \ .
\label{sig_2ph}
\end{align}

\begin{figure*}[ht]
\begin{center}
\epsfig{file=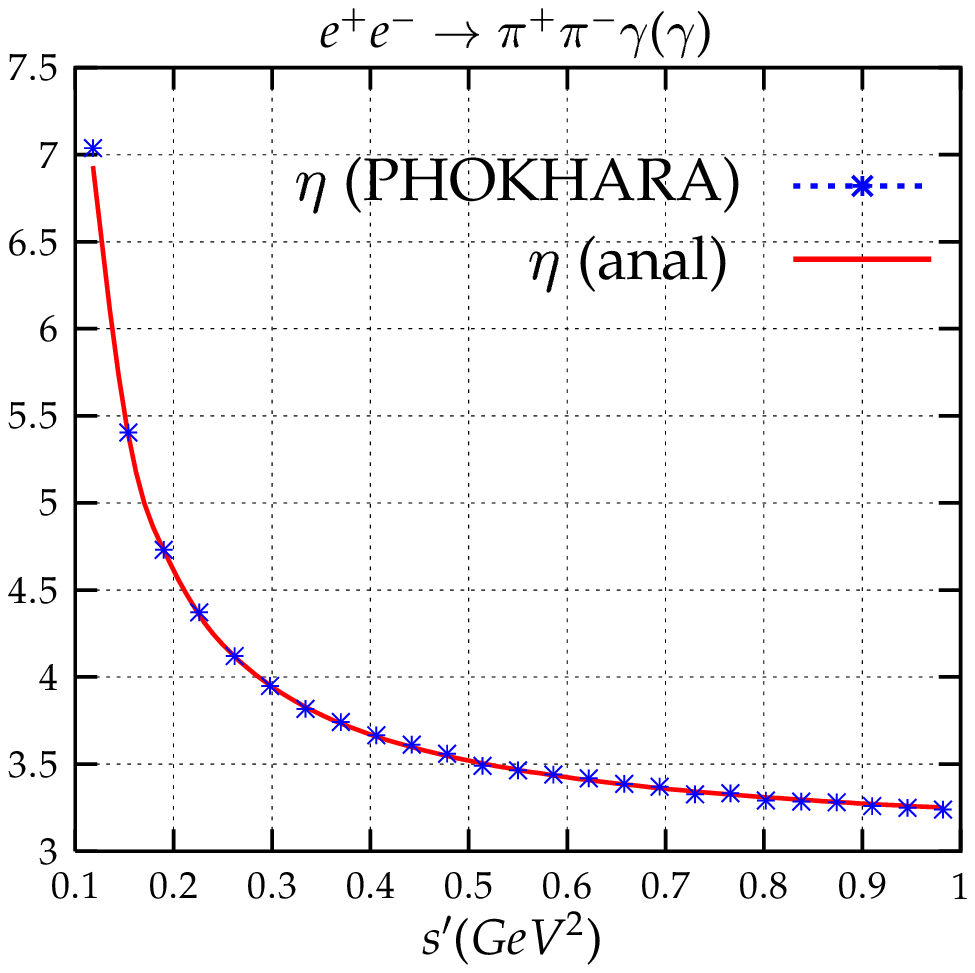,height=7cm,width=8.5cm}
\epsfig{file=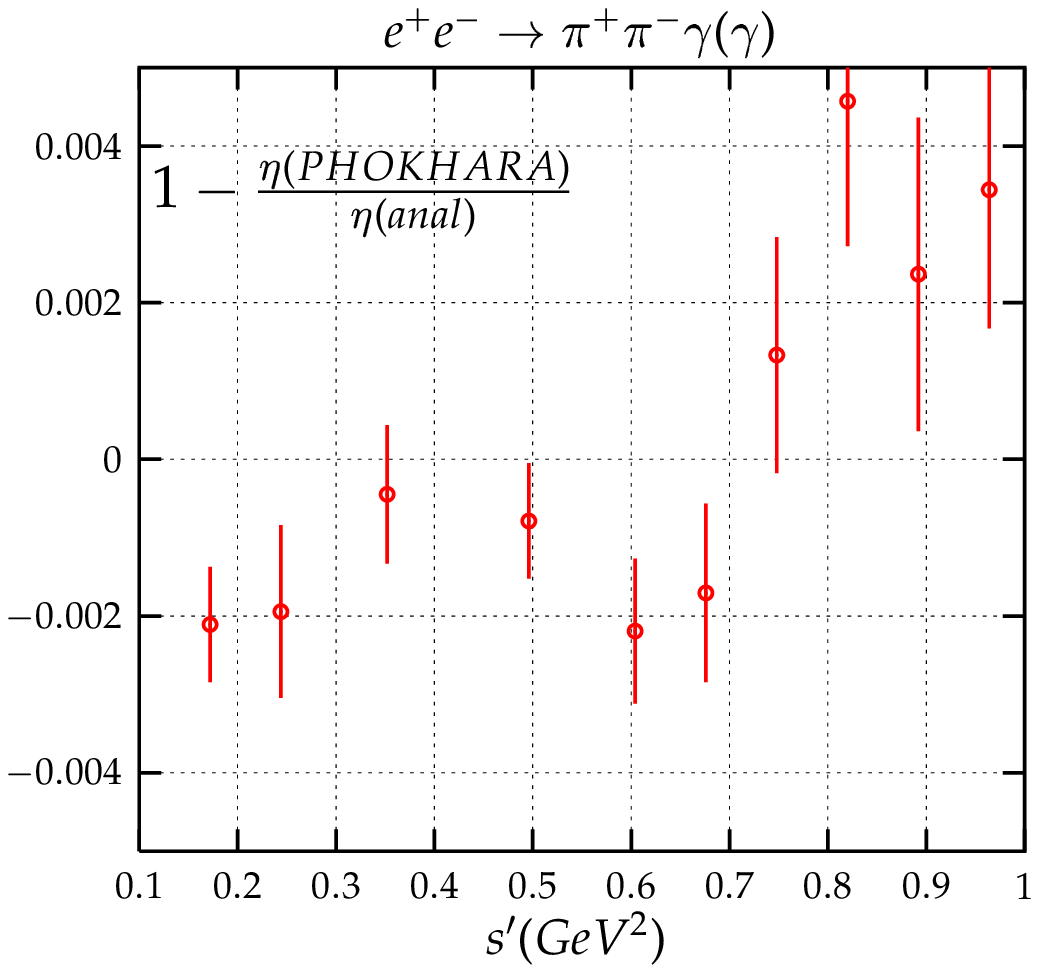,height=7cm,width=8.5cm} 
\caption{Comparison between the function \(\eta\) 
obtained from Eq.(\ref{etatot}) and from  PHO\-KHA\-RA.}
\label{fig:1-eta(ph/an)}
\end{center}
\end{figure*}

The results of the tests are shown in Fig.~\ref{fig:1-eta(ph/an)}.
The agreement of \(\eta\) calculated by PHO\-KHA\-RA and the analytical
result (\Eq{etatot}) is at the level of a few per mille.
However, since the contribution to the cross section of that part
is multiplied by \(\alpha/\pi \simeq 1/400\), the actual technical
precision of the IFSNLO PHO\-KHA\-RA cross section is
at the level of \(10^{-5}\).

\section{FSR contribution to the hadronic cross section 
and its measurement via the radiative return method}

We shall study now the impact of the new corrections on various distributions.
Before entering the discussion, let us recall the meaning of various 
abbreviations, which will be used in the following.
ISRNLO corresponds to the ISR cross section calculated at NLO without
any FSR. IFSLO includes in addition FSR at LO. Finally, IFSNLO stands for ISR 
and FSR at NLO as implemented in the new version of PHOKHARA (version 3.0).

\begin{figure}[ht]
\begin{center}
\epsfig{file=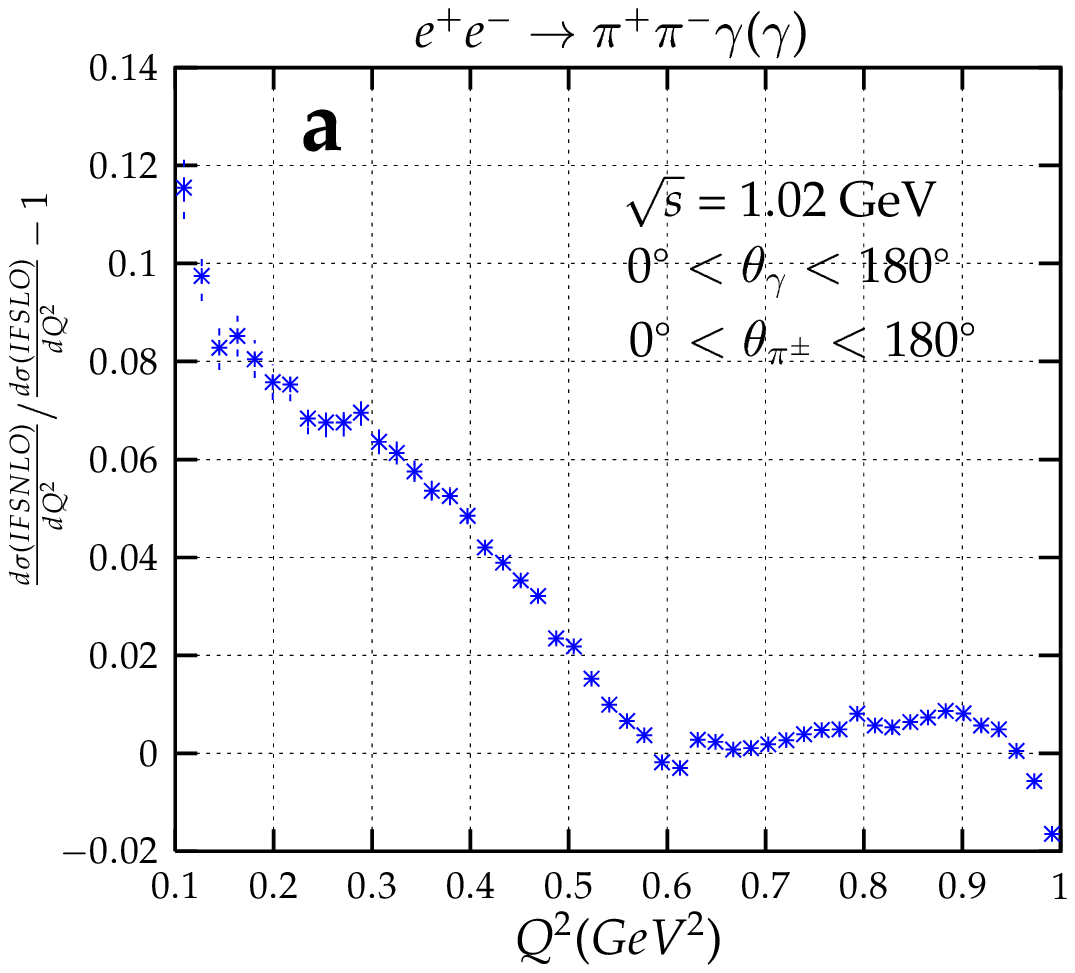,height=5.9cm,width=8.5cm}
\epsfig{file=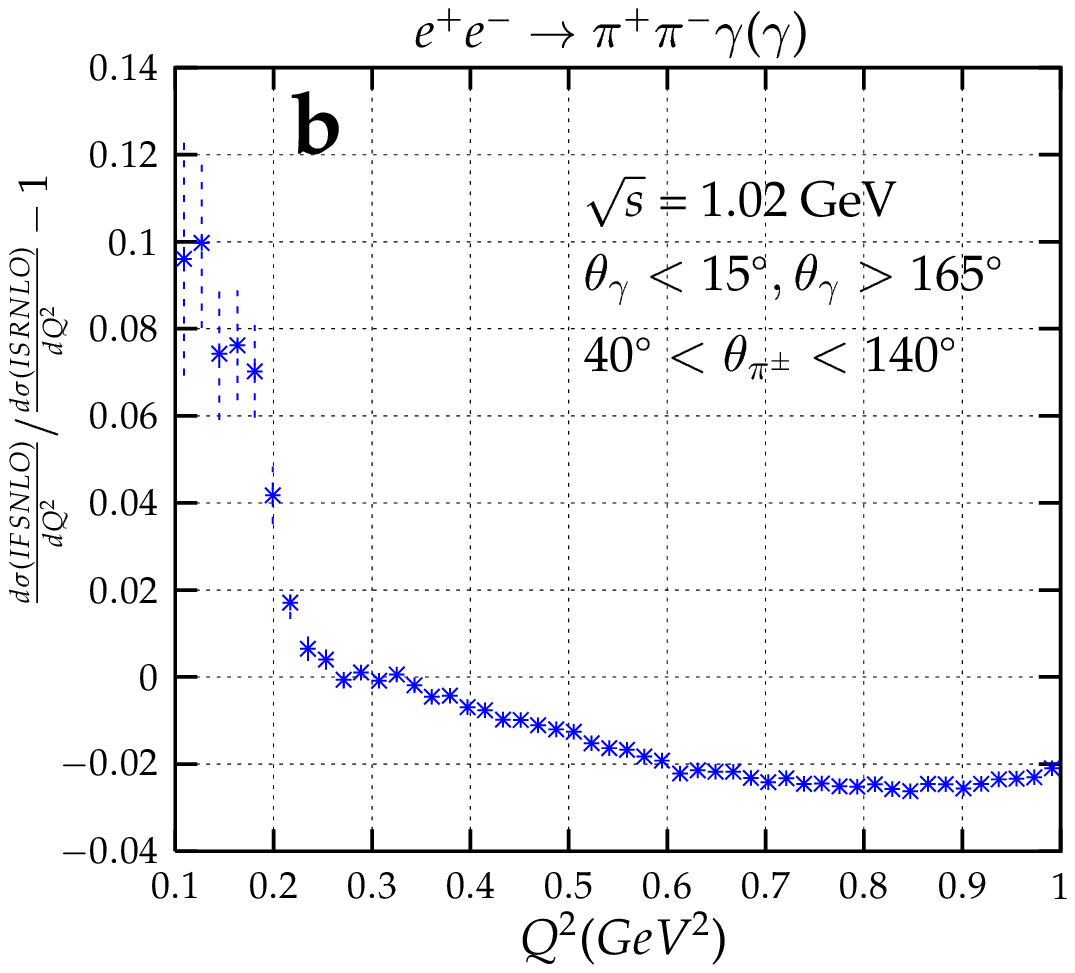,height=5.9cm,width=8.5cm}
\epsfig{file=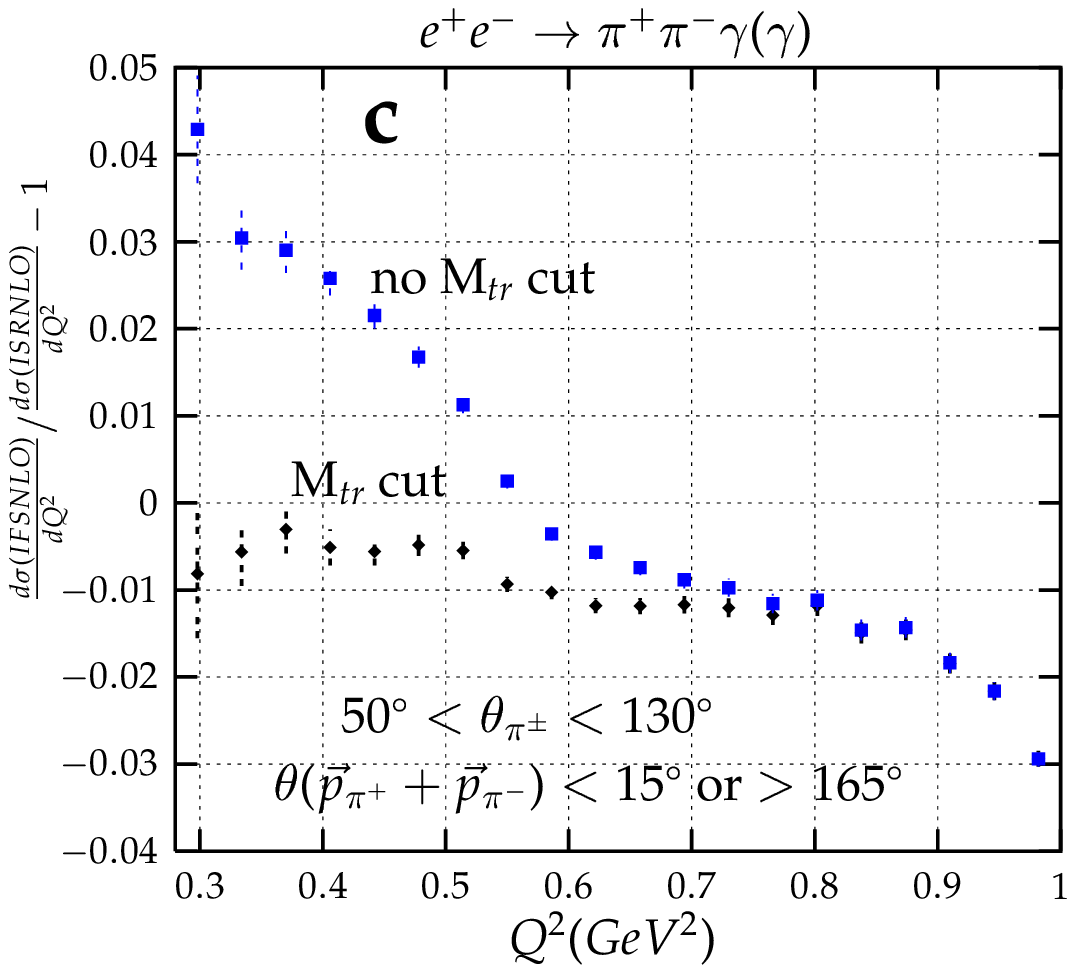,height=5.9cm,width=8.5cm}
\caption{Comparison of the \(Q^2\) differential cross sections
for \(\sqrt{s} = 1.02\)~ GeV: IFSNLO contains the complete NLO contribution, 
while IFSLO has FSR only at LO. The pion and photon(s) angles are not
restricted in (a), and restricted in (b). In (c) the cuts are imposed on
the missing momentum direction and the track mass (see text for description).}
\label{fig:ifs+lo_cut}
\end{center}
\end{figure}

Let us start the discussion for a cms energy of 1.02 GeV, relevant for the 
KLOE experiment. The IFSNLO correction to the cross section
from graphs in Fig.~{\ref{fig10}}B is relatively big
at low \(Q^2\), if no cuts are applied (Fig. \ref{fig:ifs+lo_cut}a).  
Below the \(\rho\) resonance, they grow from zero at the resonance to 
10\% of the IFSLO cross section near the production threshold, 
while they remain small above the \(\rho\) resonance. This is due to the fact
that the ISR leading to the \(\rho\) meson is strongly enhanced.
Subsequently the \(\rho\) decays into \(\pi^+\pi^-\gamma\), with a large 
contribution from the region where \(Q^2\), the
invariant mass of the \(\pi^+\pi^-\), is low. Of course
this is smeared by the width of the \(\rho\), but
the above discussion remains valid and the contribution from the 
newly implemented diagrams, through the reaction 
\(e^+e^-\to\gamma\rho(\to \gamma\pi^+\pi^-)\), is sizeable.

The effect of several `standard cuts' at \(\sqrt{s}=1.02\)~GeV
is shown in Figs.~\ref{fig:ifs+lo_cut}b and~\ref{fig:ifs+lo_cut}c. The 
sensitivity of the newly implemented contributions to these cuts can be 
exploited to test, control or experimentally eliminate FSR at NLO.
It is relatively easy to find cuts that lower the correction to 
a level of 2\%--3\%. In fact all the cuts which were used to eliminate
FSR at LO are also effective here. The standard KLOE cut on the
track mass~\cite{Achim:radcor02} reduces FSR further, to less than 1\% for most
of the \(Q^2\) range, apart of the high values of  \(Q^2\), where the
corrections remain at the level of 2\%-3\% (Fig.~\ref{fig:ifs+lo_cut}c,
lower curve).

\begin{figure*}[ht]
\begin{center}
\epsfig{file=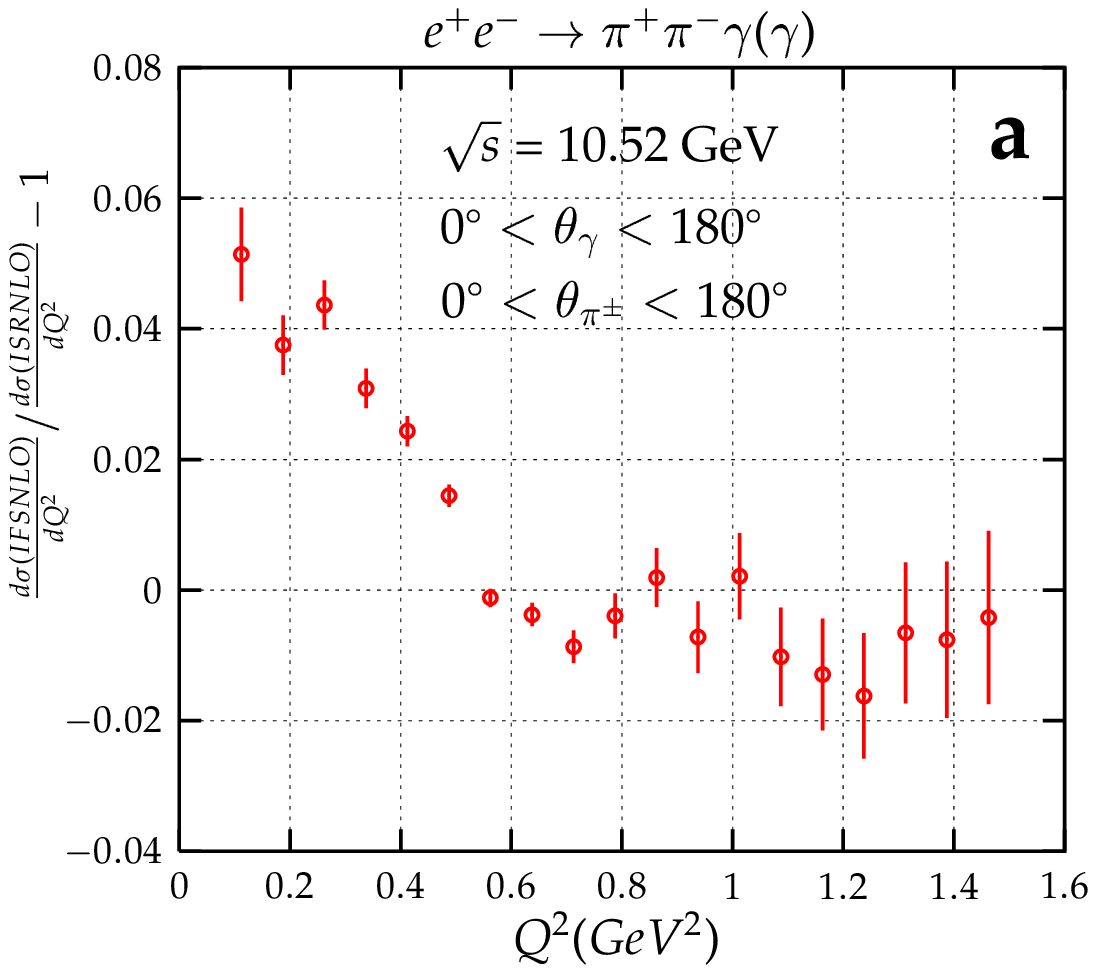,height=7cm,width=8.8cm} 
\epsfig{file=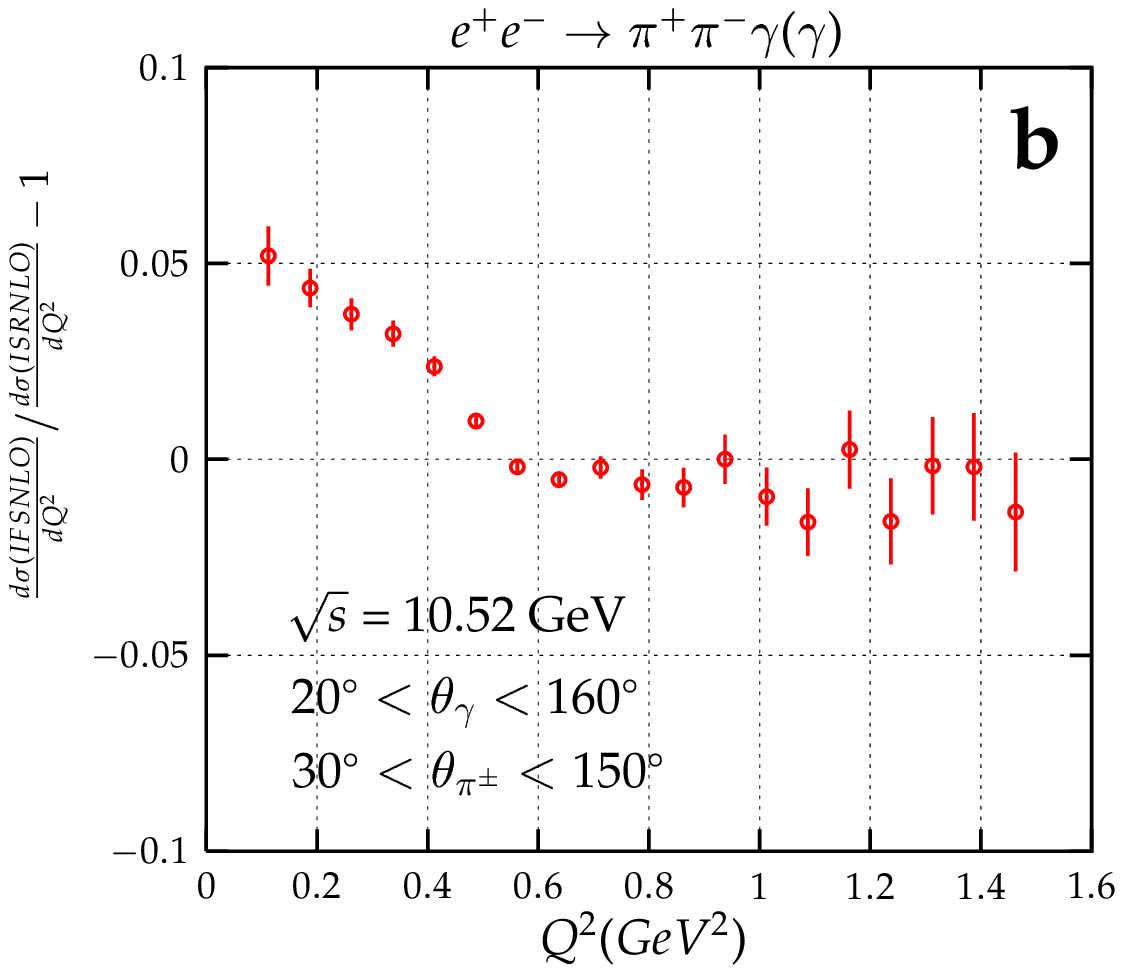,height=7cm,width=8.8cm} 
\epsfig{file=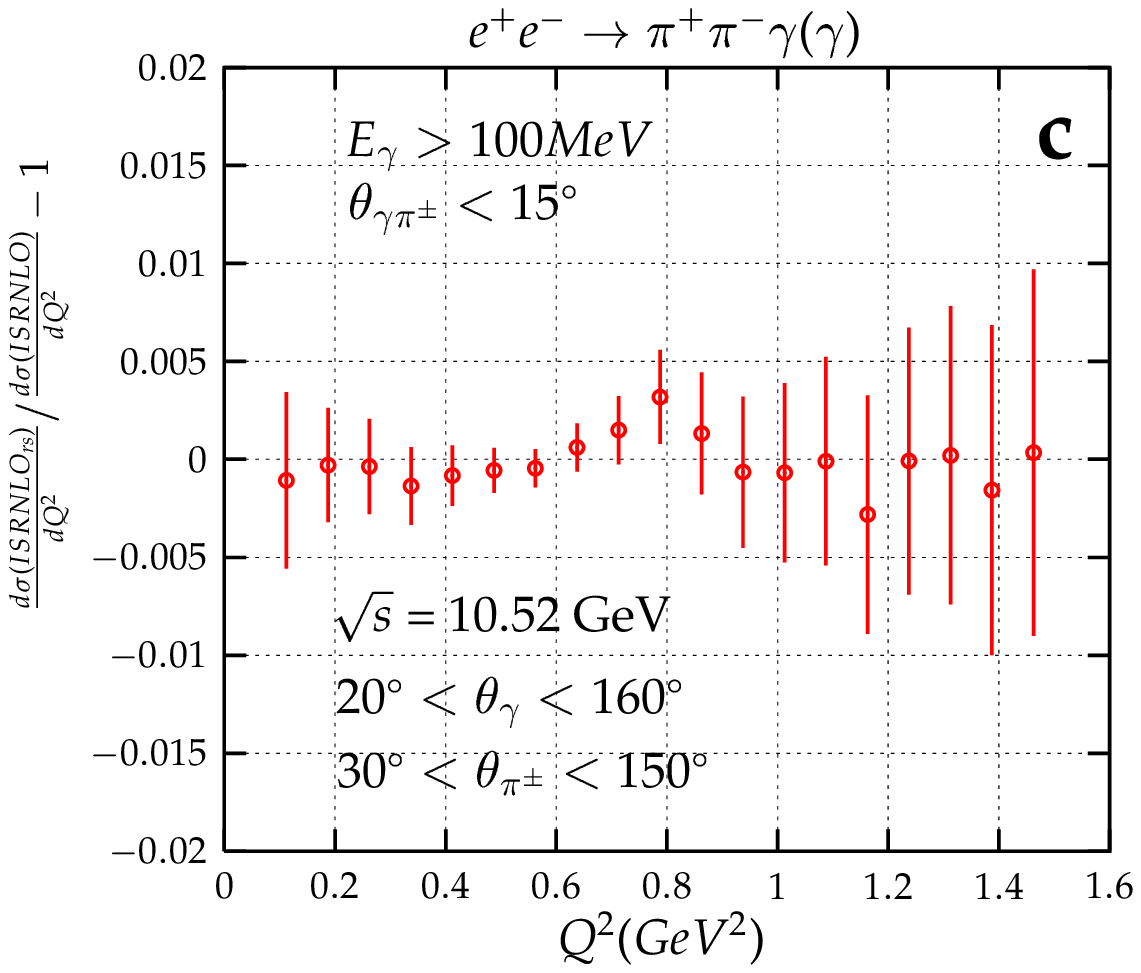,height=7cm,width=8.8cm} 
\epsfig{file=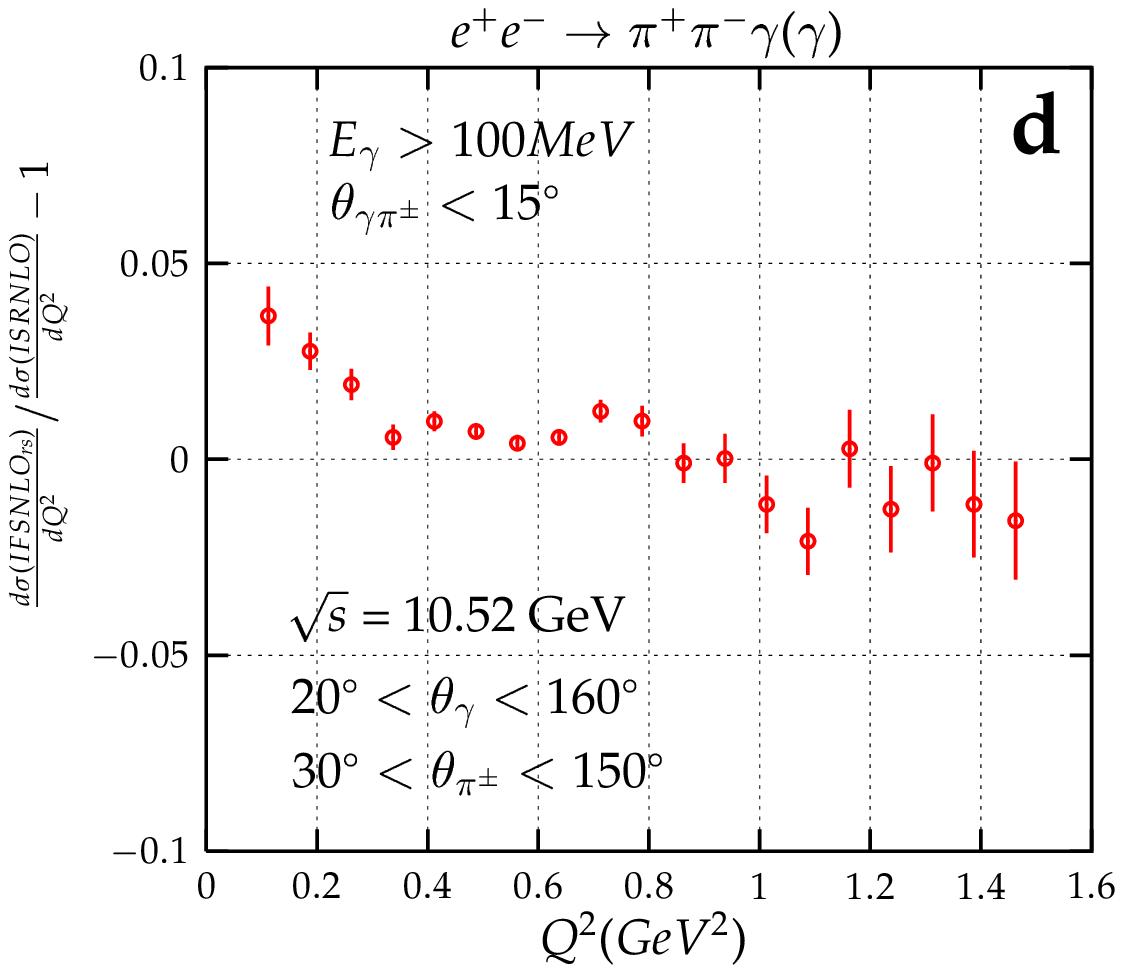,height=7cm,width=8.8cm} 
\caption{Comparisons of IFSNLO and ISRNLO distributions without (a) and with (b)
angular cuts and the effect of the `reassigning' (samples labelled $rs$)
process described in the text
on the data samples generated in ISRNLO (c) and IFSNLO (d). }
\label{fig:ifs_isr_10.52}
\end{center}
\end{figure*}

\begin{figure}[ht]
\begin{center}
\epsfig{file=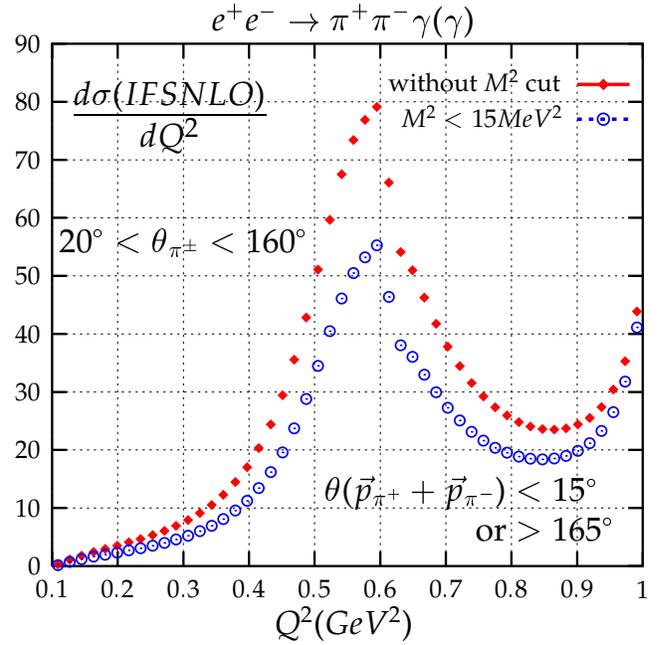,width=8.5cm} 
\caption{Dependence of the IFSNLO differential cross section
on the cut on missing invariant mass \(M^2\). }
\label{fig:fig22}
\end{center}
\end{figure}

\begin{figure}[ht]
\begin{center}
\epsfig{file=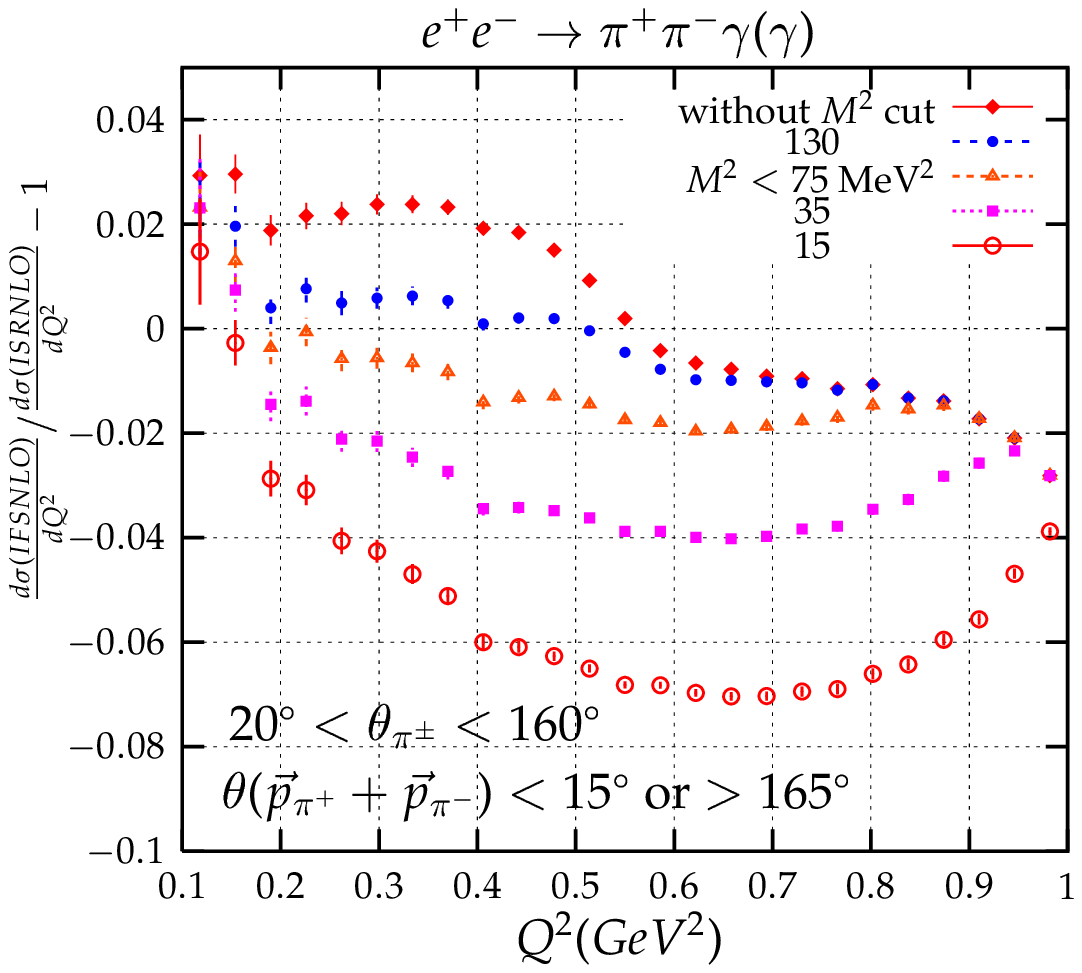,width=9cm} 
\caption{Dependence of the relative IFSNLO contribution
on the cut on missing invariant mass \(M^2\).}
\label{fig:fig23}
\end{center}
\end{figure}

The behaviour of the additional contribution at 10.52 GeV is
qualitatively similar. Without cuts the relative enhancement
amounts to approximately 4\% (Fig.~\ref{fig:ifs_isr_10.52}a),
and this remains unchanged after cuts on pion and photon 
angles, which correspond to the detector acceptance, are introduced
(Fig.~\ref{fig:ifs_isr_10.52}b). As stated before, a large fraction 
of the excess is due to the feed-down from the $\rho$-meson
in the two step process $e^+e^- \to \gamma \rho(\to \pi^+\pi^-\gamma)$.
Assuming, that hard photons, say above 100 MeV, can be identified,
we try to identify these events by assigning photons with small angles
relative to $\pi^+$ or $\pi^-$ (say 15$^\circ$) to the hadronic system when 
$M_\rho-\Gamma_\rho<\sqrt{(p_+ + p_- + p_\gamma)^2} < M_\rho+\Gamma_\rho$,
thus moving these events from $Q^2 =(p_+ + p_-)^2$ to  
$Q^2 =(p_+ + p_- + p_\gamma)^2$. For a sample generated with PHOKHARA 3.0 
without FSR at NLO the distribution remains practically the same 
(Fig.~\ref{fig:ifs_isr_10.52}c), while including FSR at NLO the excess in 
the low $Q^2$ region is significantly reduced (Fig.~\ref{fig:ifs_isr_10.52}d). 
In fact, part of the remaining enhancement (about 1 to 2\% for $Q^2$ between 
0.2 and 0.1 GeV$^2$ ) just corresponds to the effect 
of multiplication with the factor~$\frac{\alpha}{\pi}\eta(s')$.

The FSR corrections originating from the `two--step' process and
their effect on cuts necessarily introduce some model dependence
into the extraction of the pion form factor. The detection and analysis
of the complete final state --- charged pions as well as photons --- is
of course optimal for the study of this phenomena. However, the present
KLOE analysis is based on the measurement of the $\pi^+$ and $\pi^-$
momenta only. Even this partial information allows to construct
distributions, which are sensitive to final states with $\pi^+$ $\pi^-$
and two photons. As an alternative to the straightforward study 
of the \(Q^2\) distribution and its Monte Carlo simulation it is thus
even possible to test the model governing the production of $\pi^+\pi^-$
in conjunction with two photons and constrain or even measure the process 
$\gamma^* \to \pi^+\pi^-(\gamma)$, an important ingredient for the 
precise prediction of $a_\mu^\mathrm{had,LO}$.

The response of $d\sigma_\mathrm{IFSNLO}/dQ^2$ to different cuts on $M^2$,
the invariant mass of the recoiling photonic system 
(Fig.~\ref{fig:fig22}) will be an
important observable. Contributions with double emission
from the initial state and from mixed ISR--FSR will be affected by this cut.
However, the cut dependence is markedly different for $d\sigma_\mathrm{IFSNLO}/dQ^2$
(which includes both contributions) and $d\sigma_\mathrm{ISRNLO}/dQ^2$ 
(which includes configurations from ISR only) as shown 
in Fig.~\ref{fig:fig23}. The confirmation of the \(M^2\) dependence 
as predicted by PHOKHARA, as well as the verification of charge asymmetric
distributions (see Section 2) would provide additional support for the ansatz
of emission from point-like pions or allow to test alternative models.

\section{Conclusions}

Measurements of the pion form factor through the radiative return
offer the unique possibility for improved predictions for the
muon magnetic moment and the electromagnetic coupling $\alpha(M_Z)$.
Contributions to the dispersion integral from intermediate states
with $\pi^+\pi^-$ and a photon start to become relevant at the present
level of precision and can be measured in reactions leading 
to $\pi^+\pi^-$ in conjunction with two photons. An upgraded version
of PHOKHARA (version 3.0), which includes simultaneous emission
of photons from the initial and the final state, has been presented.
The two step process $e^+e^- \to \gamma \rho(\to \pi^+\pi^-\gamma)$
leads to a notable enhancement of events with low mass of the
$\pi^+\pi^-$ system. Various cuts are described, which allow to control
this effect, to identify these events, correct the distribution
and test the model for final state radiation.


\section*{Acknowledgements}

We would like to thank A.~Denig, W.~Kluge, and S.~M\"uller  for discussion of 
experimental aspects of our analysis. 
H.C. and A.G. are grateful for the support and the kind hospitality of
the Institut f{\"u}r Theoretische Teilchenphysik 
of the Universit\"at Karlsruhe.
Special thanks to Suzy Vascotto for careful proof-reading the manuscript.


\end{document}